\documentclass[
 twocolumn,
superscriptaddress,
 amsmath,amssymb,
 aps,
 pra,
 floatfix,
]{revtex4-2}
\usepackage{xcolor}
\usepackage{stackengine}
\usepackage[normalem]{ulem}
\usepackage{graphicx}
\usepackage{dcolumn}
\usepackage{bm}
\usepackage[colorlinks]{hyperref}
\hypersetup{%
	plainpages=true,
	breaklinks=true,
	hypertexnames=false,
	pageanchor=true,
	colorlinks=true,
	linkcolor={blue},
	citecolor={red},
	urlcolor={blue},
	anchorcolor={black}
}

\usepackage{physics}

\usepackage{soul}
\usepackage{accents}
\usepackage{float}
\usepackage{appendix}

\newcommand{\sz}{\hat \sigma_z}

\newcommand{\sx}{\hat \sigma_x}

\newcommand{\figref}[1]{\mbox{Fig.~\ref{#1}}}
\newcommand{\figrefs}[1]{\mbox{Fig.s~\ref{#1}}}

\newcommand{\secref}[1]{\mbox{Sec.~\ref{#1}}}

\newcommand{\appref}[1]{\mbox{Appendix~\ref{#1}}}
\renewcommand{\eqref}[1]{\mbox{Eq.~(\ref{#1})}}

\newcommand{\expec}[1]{\left\langle{#1}\right\rangle}
\newcommand{\be}{\begin{equation}}
\newcommand{\ee}{\end{equation}}
\newcommand{\bea}{\begin{eqnarray}}
\newcommand{\eea}{\end{eqnarray}}

\definecolor{darkgreen}{rgb}{0.1, 0.3, 0.0}

\newcommand{\figpanel}[2]{Fig.~\hyperref[#1]{\ref*{#1}(#2)}}
\newcommand{\figpanels}[3]{Fig.~\hyperref[#1]{\ref*{#1}(#2)-(#3)}}
\newcommand{\figpanelNoPrefix}[2]{\hyperref[#1]{\ref*{#1}(#2)}}

\usepackage{float}
\usepackage[]{changes}

\begin{document}

\title{Circuit QED Spectra
in the Ultrastrong Coupling Regime: \\
How They Differ from Cavity QED}
\author{Samuel Napoli}
\email{samuel.napoli@studenti.unime.it}
\affiliation{Dipartimento di Scienze Matematiche e Informatiche, Scienze Fisiche e  Scienze della Terra, Universit\`{a} di Messina, I-98166 Messina, Italy}
\author{Alberto Mercurio}
\affiliation{Dipartimento di Scienze Matematiche e Informatiche, Scienze Fisiche e  Scienze della Terra, Universit\`{a} di Messina, I-98166 Messina, Italy}
\affiliation{Laboratory of Theoretical Physics of Nanosystems (LTPN), Institute of Physics,
Ecole Polytechnique Fédérale de Lausanne (EPFL), CH-1015 Lausanne, Switzerland}
\affiliation{Center for Quantum Science and Engineering, EPFL, CH-1015 Lausanne, Switzerland}

\author{Daniele Lamberto}
\affiliation{Dipartimento di Scienze Matematiche e Informatiche, Scienze Fisiche e  Scienze della Terra, Universit\`{a} di Messina, I-98166 Messina, Italy}

\author{Andrea Zappalà}
\affiliation{Dipartimento di Scienze Matematiche e Informatiche, Scienze Fisiche e  Scienze della Terra, Universit\`{a} di Messina, I-98166 Messina, Italy}
\author{Omar Di Stefano}

\affiliation{Dipartimento di Scienze Matematiche e Informatiche, Scienze Fisiche e  Scienze della Terra, Universit\`{a} di Messina, I-98166 Messina, Italy}

\author{Salvatore Savasta}

\affiliation{Dipartimento di Scienze Matematiche e Informatiche, Scienze Fisiche e  Scienze della Terra, Universit\`{a} di Messina, I-98166 Messina, Italy}

\date{\today}

\begin{abstract}

Cavity quantum electrodynamics (QED) studies the interaction between resonator-confined radiation and natural atoms or other formally equivalent quantum excitations, under conditions where the quantum nature of photons is relevant.
Phenomena studied in cavity QED can also be explored  using superconducting artificial atoms and microwave photons in superconducting resonators. These circuit QED systems offer the possibility to reach the ultrastrong coupling regime with individual artificial atoms, unlike their natural counterparts. In this regime, the light-matter coupling strength reaches a considerable fraction of the bare resonance frequencies in the system.
Here, we provide a careful analysis of both incoherent and coherent spectra in circuit QED systems consisting of a flux qubit interacting with an LC resonator. Despite these systems can be effectively described by the quantum Rabi model, as the corresponding cavity QED ones, we find distinctive features, depending on how the system is coupled to the output port, which become evident in the ultrastrong coupling regime. 
\end{abstract}

\maketitle

\tableofcontents

\section{Introduction}
\label{sec:Introduction}

Superconducting quantum circuits (SQCs) based on Josephson junctions can behave like artificial atoms. They have made possible to implement atomic-physics and quantum-optics experiments on a chip \cite{Nori-atomic-physics}. In contrast to natural atoms, superconducting artificial atoms can be designed, fabricated, and controlled for various research purposes \cite{Dowling2003,10.1063/1.2155757,RevModPhys.93.025005, Vool_2017,Gu2017,Frunzio-characterizationofCQED}. Furthermore, the interaction between artificial atoms and electromagnetic fields can be controlled and engineered with more freedom \cite{RevModPhys.73.357,10.1063/1.5089550,Blais2020-oi, Devoret-howstrongcoupling}.

Superconducting artificial atoms have been used to demonstrate phenomena that cannot be realized or observed in quantum optics experiments with natural atoms. For instance, single- and two-photon processes can coexist in SQCs \cite{PhysRevLett.110.060503, Deppe2008-twoph,Qnonlinear-physreva-SavastaeKockum}, and the coupling between them and microwave fields can become ultrastrong. 
The ultrastrong coupling (USC) regime was observed for the first time in circuit QED systems in 2010 \cite{Nniemczyk2010,Forn2010}, where normalized coupling strengths  $\eta = 0.10-0.12$ were achieved. Within this platform, it is possible to explore the light-matter USC regime with a single two-level system (qubit), instead of considering many atoms or collective excitations \cite{Frisk_Kockum2019-cl,RevModPhys.91.025005, QuantumAmpl-WeiQin-Kockum-Nori,Forn-natureultrastrongcoupling-singlequbit, Bayer2017-xl, Yoshihara2017,PhysRevA.-Singlephotonfluxqubit,Gambino-Mazze-Savasta-ultrastrongnatural}.

The dimensionless parameter $\eta = g/\omega_0$, i.e., the cavity-emitter coupling rate divided by the qubit  transition frequency (or the resonance frequency of a cavity mode), is used to quantify this coupling regime. Typically, USC effects are expected when
$\eta \gtrsim 0.1$. At this value the rotating-wave approximation
(RWA) used in the weak and strong regimes starts to fail \cite{Forn-Díaz2016-brokensrule,Forn2010,Frisk_Kockum2019-cl,DeBernardis:24_Alberto_Review,Lwboitè-reviewultra} and novel physical processes can be unlocked \cite{wang2024strong,PhysRevLet-onephtwoatoms,VirtualFalci, tomonaga2024photonsimultaneouslyexcitesatoms,Qnonlinear-physreva-SavastaeKockum,Exoticquantumsate-cqed,Photon-Blockade-SavastaEridolfo}.

Recently, experiments in USC circuit QED are demonstrating a number of intriguing effects predicted theoretically \cite{wang2024strong, Wang2023}. For a correct and complete comparison between theory and data, accurate models for circuit QED systems are highly desired. 

Here, we present a theoretical framework for the calculation of emission spectra in circuit QED systems, working properly for arbitrary light-matter interaction strengths, ranging from the weak to the deep-strong coupling (DSC) regime (coupling rates larger than the bare resonance frequencies). We first study incoherent emission spectra for different coupling strengths and flux offsets under thermal-like excitation of the artificial superconducting atom. Subsequently, we inspect coherent spectra usually realized to probe circuit QED spectral proprieties. Moreover, a comparison with recent results in the literature is carried out \cite{Yoshihara2017,Phys.Rev.Semba}. Specifically, we examine a flux qubit-LC oscillator circuit where each of the two subsystems is coupled to the environment \cite{Chiorescu2004,Yan2016-or,PhysRevLett-Semba-Saito-fluxcontrol,Yoshihara2017,Yoshihara2022,Phys.Rev.Semba,PhysRevA.-Singlephotonfluxqubit}. As in usual experimental settings, we consider output photons escaping the resonator through its coupling with a coplanar open transmission line (TL). We examine two cases: (i) resonator-TL interaction through mutual inductance; (ii) capacitive resonator-TL coupling. We show that the kind of resonator-TL coupling has to be included in the model in order to provide a quantitative description of the emission process, when the system enters the USC regime. We find that the nature of such coupling can affect spectra. We also compare the numerically calculated spectra with the corresponding cavity QED ones.

\section{Theoretical Framework}

In this section, we consider a circuit QED system composed by a flux qubit galvanically coupled to an LC resonator \cite{RevModPhys.91.025005,Yoshihara2017,Yoshihara2022, Phys.Rev.Semba}. We also study the coupling of this system, through a mutual inductance $M$, to an infinite TL [see \figref{circuit}{(a)}], then the capacitive coupling of the qubit-LC circuit with a semi-infinite TL  [see \figref{circuit}{(b)}].
The full canonical quantization procedure is described in \appref{appA}.

\subsection{System Model}\label{s1}
A key feature of the flux qubit is the strong anharmonicity of its energy spectrum, hence permitting a safe \emph{two-level approximation} even in the presence of very high light-matter coupling strengths \cite{Yoshihara2022}. Thus, the Hamiltonian of the flux qubit can be easily written in the first two energy states basis ($\{\ket{g},\ket{e}\}$) as $\hat{\mathcal H}_q= \omega_0 \, \sz/2$ (henceforth $\hbar =1$), where $\omega_0=\sqrt{\Delta^2+\epsilon^2}$. Here, $\Delta$ is the tunnel splitting and $\epsilon =  2I_{p}(\Phi_{ext}- \,\Phi_{0}/2)$ the energy bias between the supercurrents flowing in opposite directions ($\{\ket{L},\ket{R}\}$), where $I_{p}$ is the critical current of the  qubit, $\Phi_{\rm ext}$ is the external magnetic flux applied to the superconducting loop, and $\Phi_{0}$ is the flux quantum \cite{10.1063/1.2155757,Yoshihara2017,Yoshihara2022,Manucharyan_2017,Chiorescu2004,PhysRevA.103.053703}. The node flux $\hat\Phi_q$ across the qubit and the coupler $L_C$ can also be projected on the $\{\ket{g},\ket{e}\}$ basis giving as result $\hat\Phi_q \simeq I_p L_C \, \hat{\Tilde{\sigma}}_x$, where $\cos{\theta}= \epsilon/\omega_0$, $\sin{\theta} = \Delta/\omega_0$, and $\hat{\Tilde\sigma}_x = (\cos{\theta}\, \hat{\sigma}_z -\sin{\theta} \,\hat{\sigma}_x)$, note that potential non-linear terms have been disregarded. Thus, the Hamiltonian $\mathcal{\hat{H}}_{0}$ of this system, describing the qubit-LC energy and their galvanic interaction, reads as  
\begin{equation}\label{qb-LC}
\mathcal{\hat{H}}_{0}= \dfrac{ \omega_0}{2} \hat{\sigma}_z+  \omega_r \hat{a}^\dagger\,\hat{a}+ \omega_r \, \eta  (\hat{a}^\dagger+\hat{a}) \hat{\Tilde{\sigma}}_x\,,
\end{equation}
where $\eta= L_C I_p \, I_{\rm zpf}/  \omega_r$ is the resonator-qubit coupling normalized with respect to the resonance frequency $\omega_r\simeq 1/\sqrt{(L_C+L)C}$ of the LC oscillator, current $I_{\rm zpf}$ and flux $\Phi_{\rm  zpf}$ zero-point fluctuations are related so that $ I_{\rm zpf}= \Phi_{\rm  zpf}/L $. It is worth stressing that whatever variation of the applied flux, which is proportional to $\epsilon$, would modify the resonance frequency of the LC circuit $\omega_r$, $ I_{ \rm zpf}$, and, as a consequence, the normalized coupling due to the presence of the coupler $L_C$.  
However, in this theoretical framework, $\omega_r$ and $ I_{\rm zpf}$ are assumed to be independent from $\epsilon$. A thorough analysis about this aspect can be found in Refs.\,\cite{Yoshihara2017,Wang2023}.

\subsection{Input-Output relations}\label{i/o_rel}
\begin{figure}[b]
    \centering
    \includegraphics[width = \linewidth]{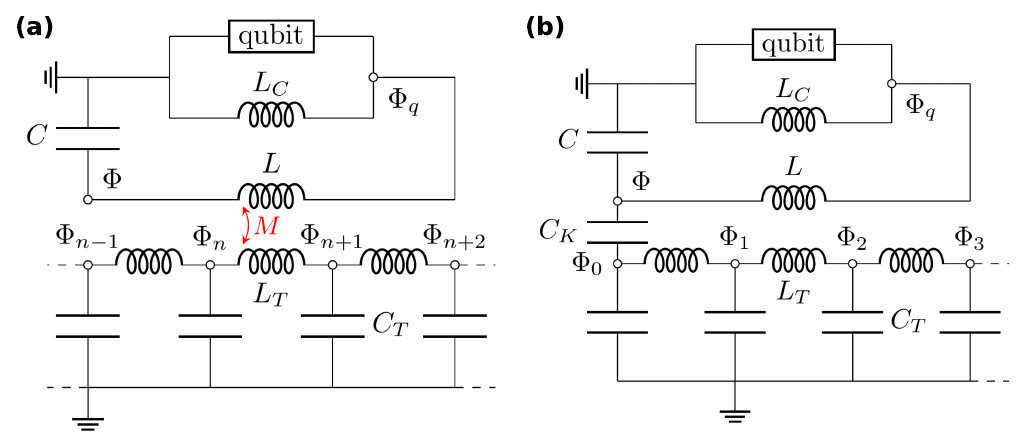}
    \caption{(a) Circuit diagram of a flux qubit galvanically coupled to the LC resonator. The infinite transmission line interacts with the inductance $L$ of the LC resonator through the mutual inductance $M$. The dynamical variables are indicated by the nodes. (b) Now, the semi-infinite transmission line is connected to the system through the capacitance $C_K$.}
    \label{circuit}
\end{figure}

In this section, we demonstrate how different coupling schemes with the TL in \figref{circuit} lead to distinct input-output relations. We will show that this aspect has consequences in the observed spectral features. 

Regarding \figref{circuit}{(a)}, the total Hamiltonian can be written as 

\begin{equation}\label{qmi}
\hat{\mathcal{H}}_{M}= \hat{\mathcal{H}}_{0}+ \hat{\mathcal{H}}_{\rm tl} + \hat{\mathcal V}_{M}\,,
\end{equation}
where $\hat{\mathcal{H}}_{0}$ is the Hamiltonian of \eqref{qb-LC}.
 Moreover, $\hat{\mathcal{H}}_{tl}=\int_{0}^{+\infty}d\omega \,  \omega \, b^\dagger_\omega b^{}_\omega$, where  $b^{}_\omega( b^\dagger_\omega)$ are the bosonic creation (annihilation) operators for the continuum frequency modes of the TL, where $\comm{b_\omega} {b^{\dagger}_{\omega'}}= \delta(\omega-\omega')$. 

The mutual interaction term results as
\begin{equation}\label{vind}
\hat{\mathcal V}_{M}= i\, \frac{1}{\Phi_{ \rm zpf}} \, \hat{\Phi}_L \int_{0}^{+\infty} d\omega\, \,g(\omega)  \, [\hat{b}_\omega-\hat{b}^\dagger_\omega ]\,,
\end{equation}
where $\hat{\Phi}_L= \hat \Phi - \hat \Phi_q = \Phi_{\rm zpf}( \hat a + \hat{a}^\dagger-2\eta\, \hat{\tilde{\sigma}}_x)$ and $g(\omega) = \sqrt{\gamma_r \,\omega/(2 \pi \omega_r)}$, with $\sqrt{\gamma_r/(2\pi)}  = \alpha \Phi_{ \rm zpf}\, (\Lambda/v_0) \sqrt{\omega\, \omega_r}$. Furthermore, $\Lambda=\sqrt{{Z_0}/{4\pi}}$, $Z_0$ is the TL impedance, $\alpha=M/L\,l_T$, and $v_0= 1/\sqrt{l_T\,c_T}$ is the velocity of the light in the TL, $l_T= L_T/\Delta x$ and $c_T= C_T/\Delta x$ are the inductance and the capacitance at each site of the TL per unit of length, $\Delta x$ is the distance between two sites.

Figure~\ref{circuit}(b) displays the flux qubit-LC-oscillator system capacitively coupled to the TL. We will show that the different coupling mechanisms in \figrefs{circuit}{(a)} and (b) can influence the observed spectra, especially when the coupling between the flux qubit and the LC oscillator reaches the USC or DSC regime.
The total Hamiltonian is 
\begin{equation} \label{capfull}
\hat{\mathcal{H}}_{C}= \hat{\mathcal{H}}_{0}+ \hat{\mathcal{H}}_{ \rm tl} + \hat{\mathcal V}_{C}\, ,
\end{equation}
where the interaction with the TL reads as
\begin{equation}\label{vcap}
\hat{\mathcal V}_{C} = \; i (\hat a^\dagger- \hat a)\,   \int_{0}^{+\infty} d\omega \, g(\omega)\,  i(\hat b^{\dagger}_{\omega}-\hat b_{\omega})\,,  
\end{equation}
where, in this case, $g(\omega) =\sqrt{\gamma_r\,\omega/(2 \pi\,\omega_r)}$ and ${\sqrt{\gamma_r} = C_K\, \omega_r/\sqrt{ v_0 c_T C}}$.
We observe that the system operator $\hat X_{C}= i (\hat a^\dagger- \hat a)$, appearing in the system-TL interaction Hamiltonian [see \eqref{vcap}], differs from that in the case of inductive coupling: $\hat X_{M}=  \hat a + \hat{a}^\dagger-2\eta\, \hat{\tilde{\sigma}}_x$. This difference, as expected, has a direct impact on the input-output voltage expressions. In \appref{appIO}, we provide the derivation for both cases in detail.

We find that for the mutual inductive coupling 
\begin{equation}\label{VM}
\hat{V}^{\pm}_{M_{\rm out}}(x, t) =  \hat{V}^{\pm}_{\rm in} (x,t)-\,\frac{ M}{2  L }   \, \dot{\hat{\Phi}}^{\pm}_L(x,t)\,, 
\end{equation}
and for the capacitive coupling
\begin{equation}\label{VC}
\hat{V}^{\pm}_{C_{\rm out}}(t - x/v_0) = \hat{V}^{\pm}_{in}(t+x/v_0) + Z_0 \frac{C_K}{C} \dot{\hat{Q}}^{\pm}(t - x/v_0)\,.
\end{equation}
Here, the $\pm$ label indicates positive $(+)$ and negative $(-)$ frequency operators.

\subsection{Master Equation and Emission Spectra}\label{mees}

In the following, we will diagonalize numerically the closed system Hamiltonian $\mathcal{\hat H}_0$ and will treat the interaction of the system with the external reservoirs, including the input-output TL, within a master equation approach suitable for light-matter systems in the USC regime. 

Both \eqref{qmi} and \eqref{capfull} lead to generalized master equations (GMEs), as discussed in Ref.\,\cite{PhysRevA.98.053834}:
\begin{equation}\label{lou}
\dot{\hat \rho} = - i \comm{\mathcal{\hat H}_0}{\hat \rho}+\mathcal{L}_{g} \, \hat \rho\,.
\end{equation}
The Liouvillian superoperator $\mathcal{L}_{g} $ involves the operators that are responsible for the interaction of the system with the baths (the complete equation is presented in \appref{appB}). The TL, schematically displayed in \figref{circuit}, is one of the system baths. Additional bath-channels can, for example, describe qubit spontaneous emission losses and internal losses of the resonator.
We also observe that the two different schemes in \figrefs{circuit}{(a)} and (b) give rise to different terms in the Liouvillian superoperator (see \appref{appB}).
Regarding the qubit, we assume that the interaction with the relative reservoir occurs through the operator $\hat{\tilde{\sigma}}_x$ for both cases (the same operator in the qubit-LC interaction term).
Also, the damping rates of the system are reported in \appref{appB}.

In \secref{emissionspectra}, we present emission spectra under incoherent (thermal-like) qubit excitation, in the absence of input drive $ \expec{\hat{V}^{\pm}_{in}}\approx 0$. 
As a consequence, for the configuration in \figref{circuit}(a), the voltage measured from the input-output port is determined by the following system operator [see \eqref{VM}]
\begin{equation}
\dot{\hat{\Phi}}_L \propto
\dot{\hat X}_{M} =  \bigg[i\,(\hat{a}^\dagger-\hat{a})
-2\eta\,\frac{\omega_0}{\omega_r} \sin{\theta}\,\hat{\sigma}_y\bigg]\,,
\end{equation}
for the configuration displayed in \figref{circuit}(b), the output voltage is determined by [see \eqref{VC}]
\begin{equation}
\dot{\hat{Q}} \propto \dot{\hat X}_{C} =   -(\hat{a}^\dagger+\hat{a}+2\,\eta \, \hat{\tilde{\sigma} }_x)\,. 
\label{Qdot}\end{equation}

Once the density matrix at the steady state $\hat{\rho}_{ss}$ is numerically derived from \eqref{lou}, and applying the quantum regression theorem \cite{gardiner2004quantum, PhysRevResearch.4.023048}, the power spectrum can be expressed as 
\begin{equation}\label{emission}
S_{M(C)}(\omega)\propto \Re\int_{0}^{+ \infty}\!\! \! \! \!\! \! d\tau  e^{-i \omega \tau} \expec{\dot{\hat X}_{M(C)} ^{(-)}(t+\tau) \dot{{\hat X}}_{M(C)} ^{(+)}(t)}_{ss},
\end{equation}
where, for a generic system operator $\hat S$,
\begin{equation}
\hat{S}^{(+)}=\sum_{i,j>i}  {S}_{ij}\,\ketbra{i}{j}\,  
\end{equation}  
represents the positive frequency component of $\hat S$,
$\{\ket{i}\}$ are the eigenstates of $\hat{\mathcal{H}}_{0}$ ordered according to growing energies. The negative frequency component $\hat{S}^{(-)}$ also corresponds to $(\hat{S}^{(+)})^{\dagger}$.
 All the remaining constant factors deriving from the \eqref{VM} and \eqref{VC} are not considered. Moreover, the time shift between the negative and positive frequency components and the spatial dependence have no impact on the resulting spectra, since we evaluate the power spectra at the steady state.

\subsection{Reflectivity}\label{cohspect}

In this section, we first describe the method used to compute the coherent (reflectivity) spectra. In \secref{cohspectra}, we present the numerical results, setting the parameters to match those corresponding to a specific circuit QED system analyzed in Refs.\,\cite{Yoshihara2017,Phys.Rev.Semba}.

Once again, we find that in the USC or DSC regime, certain features of the spectra depend on the specific physical observable which is measured. 

As already seen, the RWA fails in the USC regime, as a consequence,
the time-dependence of the coherent drive (input signal) cannot be removed, and it is necessary to go beyond standard input-ouput relations \cite{gardiner2004quantum,Walls2008}.
Moreover, the time-dependence of the driving term in the Hamiltonian prevents the possibility to obtain a stationary steady-state density matrix. The steady state density matrix, $\hat{\rho}_{ss}(t)$, has now a periodic time-dependence  ($T=2\pi/\omega_d$; $\omega_d$ is the frequency of the input signal). This corresponds to the Floquet formalism, as pointed out in Refs.\,\cite{wang2024strong,Wang2023,Raman-Alberto,SalmonGustinSettineriDiStefanoZuecoSavastaNoriHughees-guagefreedomspectra}. Hence, the steady state can be expanded as $\hat{\rho}_{ss}(t)= \sum_{k = -\infty}^{+\infty}\hat \rho^{k}\, e^{i\,k\omega_d\,t}$, where $k$ is an integer number. Using this relation and the generalized master equation (this matter is discussed in \appref{appB}), the various components of the density matrix can be obtained and the $k = -1$ term is used to calculate linear coherent spectra. 
Specifically, when computing the expectation value of a positive frequency operator, e.g., $\langle \hat X^+ \rangle(\omega_d)$ at the frequency of the drive, the contribution from $k=-1$ is the only relevant term.

Considering a driving tone in a coherent state
\begin{equation}
\expec{V^{+}_{\rm in}}(t) = -i \abs{b_{\rm in}} \Lambda \sqrt{\omega_d} e^{-i \omega_d \,t}\,,
\end{equation}
where $\abs{b_{\rm in}}^2$ is the photon rate of the coherent drive,
the frequency-dependent reflectivity $S^M_{11}$ for the mutual inductive coupling between the system and the TL  can be directly derived from \eqref{VM}, resulting in
\begin{equation}\label{transm}
S^M_{11} =\abs{\frac{\expec{\hat{V}^{+}_{M_{\rm out}}}}{\expec{V^{+}_{\rm in}}}}=\abs{1- \frac{\sqrt{2\pi}}{\abs{b_{\rm in}}} \sqrt{\frac{\omega_d \, \gamma_{m}}{\omega_r}}\, \Tr{\hat X^{+}_{M}\,\hat \rho^{-1}}}\, .
\end{equation}
In \eqref{transm}, $\gamma_{m}$ represents the damping rate of the input-output port. Furthermore, the spatial dependence in the $S^M_{11}$ derivation is neglected, since we will evaluate the spectra at the steady state and far away from the point of interaction between the circuit QED system and the TL. 
We also assumed here $\dot {\hat X}^{+}_{M} \approx -i\omega_d \, \hat X^{+}_{M}$ at the steady state. 

In \secref{cohspectra}, we compare our spectra with those in Refs.\,\cite{Yoshihara2017,Phys.Rev.Semba}. In that case, $\hat X^{+}_{M}$ will be replaced with $(\hat a+ \hat a^\dagger)^{+}$.
Notice that \eqref{transm} describes a one-port excitation-detection configuration from the point of view of the circuit QED system (reflectivity). However, this configuration could also be interpreted as a two-port scheme from the point of view of the TL, since the signal is detected from the opposite side with respect to the input.

Similarly, starting from \eqref{VC}, the reflectivity in the capacitive coupling scheme is written as 
\begin{equation}\label{reflect}
S^C_{11} =\abs{\frac{\expec{\hat{V}^{+}_{C_{\rm out}}}}{\expec{V^{+}_{\rm in}}}}=\abs{1 + \frac{\sqrt{2\pi}}{\abs{b_{\rm in}}} \sqrt{\frac{\omega_d \, \gamma_{c}}{\omega_r}}\, \Tr{\hat X^{+}_{C}\,\hat \rho^{-1}}}\,, 
\end{equation}
where $\gamma_{\rm c}$ is the damping rate of the input-output port.

\section{Emission Spectra}\label{emissionspectra}

In this section, we investigate the emission properties of the system under incoherent excitation of the qubit. For the sake of simplicity, we assume a zero temperature for the photonic reservoir ($T_r =0$). Moreover, we consider the following damping rates:  $\gamma_q/\Delta = 10\; \gamma_r/\omega_r= 10^{-2}$, so the system losses originate mainly from the qubit. The incoherent excitation of the qubit is modeled by assuming a qubit reservoir at an effective temperature $T_q/\omega_r \neq 0$ (the Boltzmann constant $k_B$ is put to 1). The numerical simulations are performed with QuantumToolbox \cite{quantumtoolboxjl}, which is a cutting-edge Julia package designed for quantum physics simulations, closely emulating the popular Python QuTiP package \cite{JOHANSSON20121760,JOHANSSON20131234}.

\subsection{Parity symmetry (zero flux-offset)}\label{zeroep}
\begin{figure}[t]
    \centering
    \includegraphics[width = 0.5\textwidth]{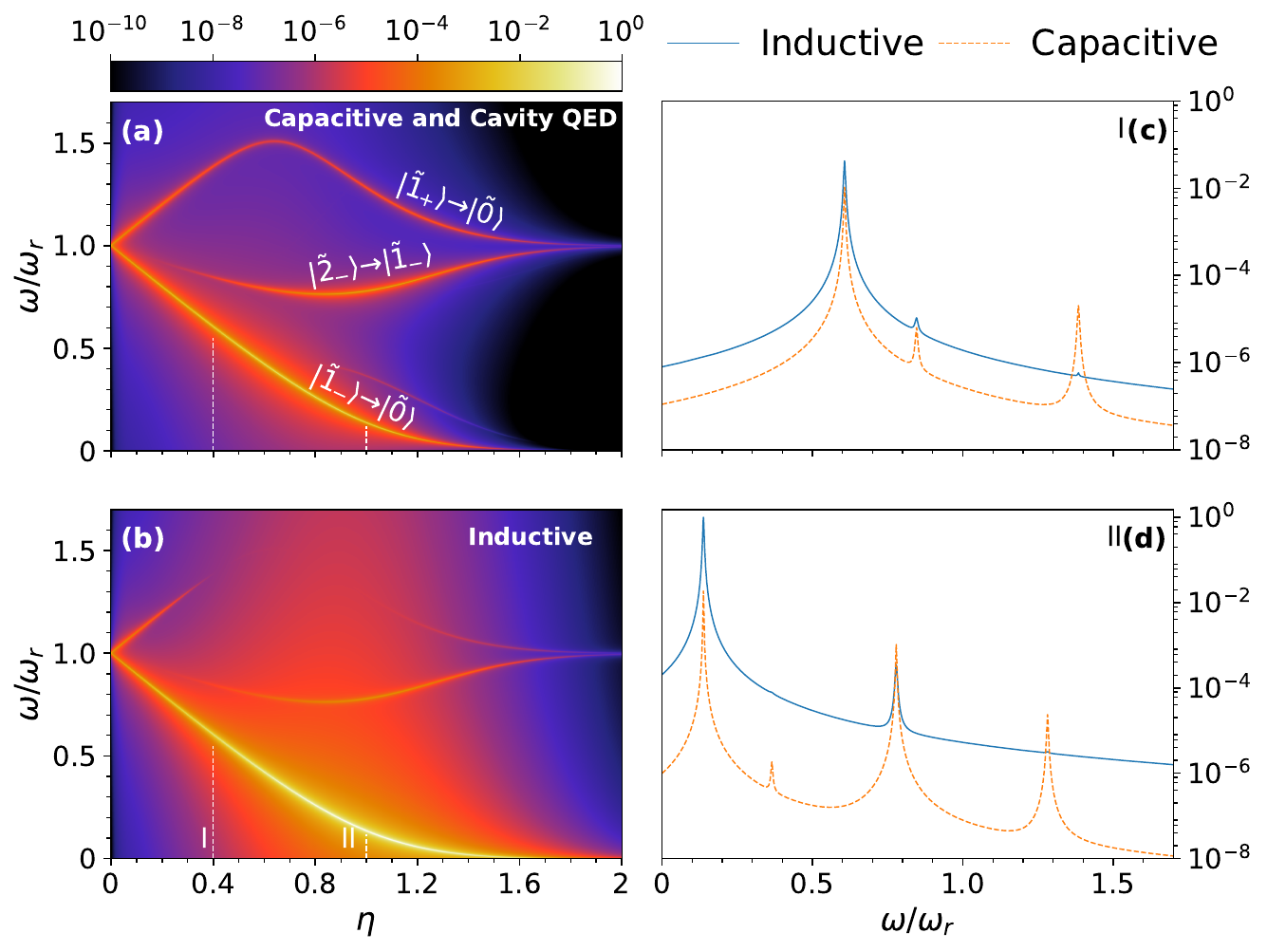}
    \caption{Logarithmic emission spectra obtained for $T_q/\omega_r= 0.1$, $\Delta/\omega_r = 1$, and $\epsilon = 0$. (a) Logarithmic spectra $S_C(\omega)$ as a function of $\eta$ at the steady state $\hat \rho_{ss}$, for the system capacitively coupled to the TL (also valid for the cavity QED case). (b) Logarithmic spectra for the mutual inductive coupling,  $S_M(\omega)$, with the TL. Spectra (a) and (b) are normalized with respect to the absolute maximum value. (c) Logarithmic spectrum computed for $\eta=0.4$. (d) In this case, the logarithmic spectrum is evaluated at $\eta= 1$. Spectra (c) and (d) are normalized with respect to the maximum value of spectra (a) and (b).}
    \label{fig1}
\end{figure}
We start analyzing the zero-flux offset case ($\epsilon=0$, implying $\theta = \pi/2$), corresponding to a system described by the standard quantum Rabi model (QRM) Hamiltonian [see \eqref{qb-LC}].

We study the non-equilibrium dissipative dynamics of this circuit-QED system under incoherent (thermal-like) excitation.
In particular, we solve numerically the steady-state master equation, setting the qubit effective-temperature $T_q/\omega_r=0.1$.
The system incoherent excitation through the qubit reservoir is able to populate the lowest energy excited states of the system $\ket{\tilde n_{\pm}}$, which in turn decay towards the lower energy states. In the QRM, beyond the strong-coupling regime, the eigenstates do not exhibit the simple structure of the Jaynes-Cummings (JC) model. Therefore, we use a generalized notation for these states by introducing a tilde \cite{PhysRevA.84.043832, PhysRevResearch.4.023048}. In particular, the system energy states are labelled so that in the small $\eta$ limit $\ket{\tilde n_{\pm}}$ coincides with the corresponding JC state $\ket{n_{\pm}}$.

\begin{figure}[t]
    \centering
    \includegraphics[width= 1 \linewidth]{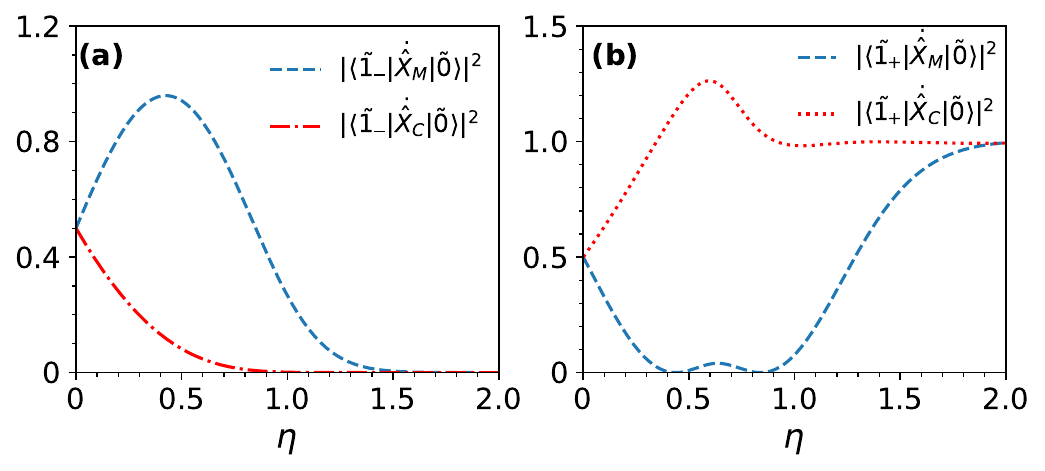}
    \caption{Modulus squared of two transition matrix elements as a function of $\eta$, where $\epsilon=0$ and $\Delta/\omega_r=1$. Panel (a) shows $\abs{\bra{\tilde1_-}\dot{\hat{X}}_{M(C)}\ket{\tilde 0}}^2$, while panel (b) displays $\abs{\bra{\tilde1_+}\dot{\hat{X}}_{M(C)}\ket{\tilde 0}}^2$.  } 
    \label{m1}
\end{figure}

Figures~\ref{fig1}(a) and (b) present the resulting emission steady-state spectra as a function of the system coupling rate $\eta$. Specifically, \figref{fig1}(a) shows the case of capacitive coupling with the TL, while \figref{fig1}(b) displays the spectra obtained for the inductive coupling with the input-output TL. Both panels clearly highlight the transitions $(\tilde 1_-, \tilde 0)$, $(\tilde 2_-,\tilde 1_-)$, and $(\tilde 1_+,\tilde 0)$ (in ascending order of frequency). As expected, the $(\tilde 1_-, \tilde 0)$ transition line is the most intense for any values of $\eta$ at such low effective temperature, being the lowest energy transition. The other transitions tend to become brighter at increasing coupling strength, until the decoupling fate occurs in the DSC limit \cite{Decoupling-deliberato,Salado-Mejía_2021_thermodynamicsultrastrong}.  We observe that the mutual inductive spectra exhibit a quenching of the $(\tilde 1_+,\tilde 0)$ spectral line when the coupling ranges roughly from 0.4 to 1 [see \figref{fig1}(b)]. This different behaviour between the two spectra is determined by the matrix elements of the corresponding system observables coupled to the output channels (see \figref{m1}). Figures~\ref{fig1}(c) and (d) display the spectra obtained  fixing specific values of the normalized coupling: $\eta =0.4,\, 1$; respectively. 

Figure \ref{m1} shows the modulus squared of the system matrix elements determining the output emission associated to the transitions $ \ket{\tilde 1_-} \to | \tilde 0 \rangle$ and $| \tilde 1_+ \rangle \to | \tilde 0 \rangle$, versus the normalized coupling strength $\eta$: $\abs{\bra{\tilde1_-}\dot{\hat{X}}_{M(C)}\ket{\tilde 0}}^2$  [\figref{m1}(a)] and $\abs{\bra{\tilde1_+}\dot{\hat{X}}_{M(C)}\ket{\tilde 0}}^2$ [\figref{m1}(b)].

These results are a clear evidence how the calculated spectra are influenced by the observable coupled to the output port. This influence, however, becomes evident only in the USC or DSC regime. It is also interesting to compare these findings for circuit QED systems with the corresponding results for a cavity QED model, we consider the atomic transition interacting with the cavity field via standard dipolar coupling \cite{Babiker-Loudon-Gauge,CohenTannoudji1997, DiStefano2019,Gauge-inviariant-Hughes,Settineri_2021,PhysRevResearch.4.023048}, and where the photodetection rate is proportional to the expectation value of the product of the negative and positive electric-field operators \cite{GlauberPhysRev}: $\hat{X}_{D} = i (\hat{a}-\hat{a}^\dagger)-2\eta\, {\sx}$. It turns out that the dipolar cavity QED spectra coincide with the circuit QED ones computed for the capacitive coupling to the output TL [see \figref{fig1}(a)]. Looking at the pertinent equations, this can be understood by observing that, applying a phase rotation $\hat a \to i \hat a$ ($\hat a^\dag \to -i \hat a^\dag$) to the cavity QED Hamiltonian and to the electric-field operator, the circuit QED Hamiltonian, and the output operator proportional to $\dot{\hat Q}$ [see \eqref{Qdot}] are obtained. We can interpret this correspondence, noting that the output capacitive coupling carries on the information of the electric field.

In \appref{appspec}, we report additional parity-symmetry spectra at a higher effective temperature $T_q$.

\begin{figure}[b]
    \centering
    \includegraphics[width = 0.5 \textwidth]{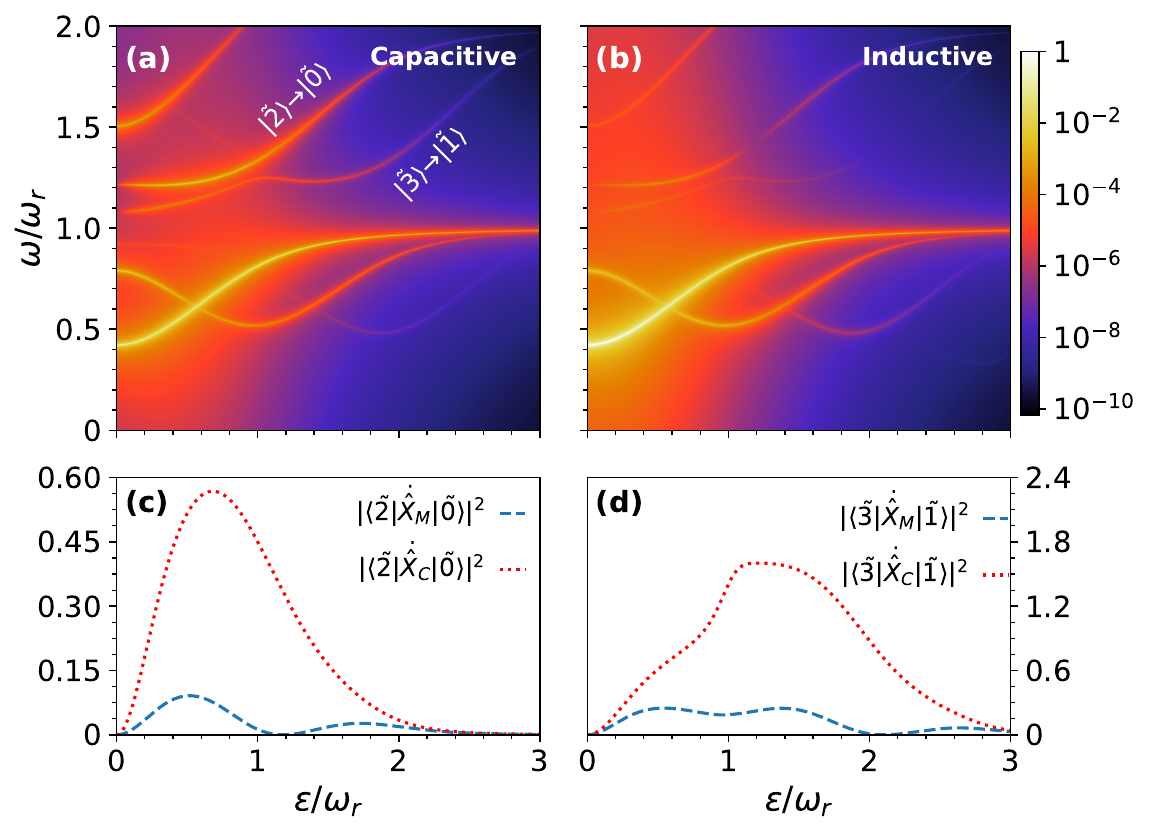}
    \caption{Logarithmic emission spectra obtained for $T_q/\omega_r= 0.15$, $\Delta/\omega_r = 1$, and $\eta = 0.6$. (a) Logarithmic spectra $S_C(\omega)$ as a function of $\epsilon/\omega_r$, for the system capacitively coupled to the TL. (b) Logarithmic spectra for the mutual inductive coupling,  $S_M(\omega)$, with the TL. Spectra (a) and (b) are normalized with respect to the absolute maximum value. The specific spectral lines, discussed in the text, are indicated in the panel (a). Panels (c) and (d) report two relevant transition matrix elements of $\dot{\hat X}_C$ and $\dot{\hat X}_M$ (the transitions displaying differences, see text).}
    \label{2parte}
\end{figure}


\subsection{Symmetry breaking (non-zero flux offset)}\label{ediv0}

Circuit QED systems consisting, for example, of an LC oscillator interacting with a flux qubit are usually characterized by detecting spectra as a function of the qubit flux offset $\epsilon$ \cite{Yoshihara2017,Yoshihara2022,Phys.Rev.Semba}. For $\epsilon \not = 0$, the parity selection rule of the standard QRM breaks down and the system can be described by a generalized quantum Rabi model [see \eqref{qb-LC}]. In this subsection, we present incoherent emission spectra versus the flux offset $\epsilon$ for a normalized coupling strength $\eta = 0.6$. We also consider an effective temperature of the qubit $T_q/\omega_r = 0.15$. Figure \ref{2parte} displays the evolution (as a function of the flux offset) of several transitions. Note that we have switched the notation of the energy eigenstates from $\ket{\tilde n_{\pm}}$ to $\ket{\tilde n}$, increasing $n$, the energy grows.

We observe that the lowest transition, $(\tilde 1 , \tilde 0)$, is the most dominant for any $\epsilon$ for both the interaction kinds with the TL.

Figures \ref{2parte}(c) and (d) show the modulus squared of the expectation values of the relevant output system operators determining the intensity of two  emission lines (associated to the transitions $| \tilde 2 \rangle \to | 0 \rangle$ and $| \tilde 3 \rangle \to | 1 \rangle$ ). As explicitly illustrated in \figrefs{2parte}(c) and (d), these two transitions are forbidden at $\epsilon =0$. Moreover, we observe that these matrix elements explain the different intensities of the lines highlighted in \figref{2parte}(a) and \figref{2parte}(b).

In \appref{appspec}, we investigate emission spectra in the DSC regime and also varying $\Delta$.

\section{Coherent Spectra}\label{cohspectra}
In this section, we examine the coherent emission properties of the system.
Specifically, we calculate reflectivity spectra obtained coupling inductively [$S^M_{11}(\omega_d)$] and capacitively [$S^C_{11}(\omega_d)$] the circuit-QED system with an input-output TL, according to the configurations sketched in \figref{circuit}.
For the calculations, we considered parameters which are very similar to those describing
 \'\,CIRCUIT III\,\' \, in Ref.\,\cite{Yoshihara2017}: $\eta = 1.01$ and $\Delta/\omega_r= 0.69$.
\begin{figure}[htb]
    \centering
    \includegraphics[width= 1 \linewidth]{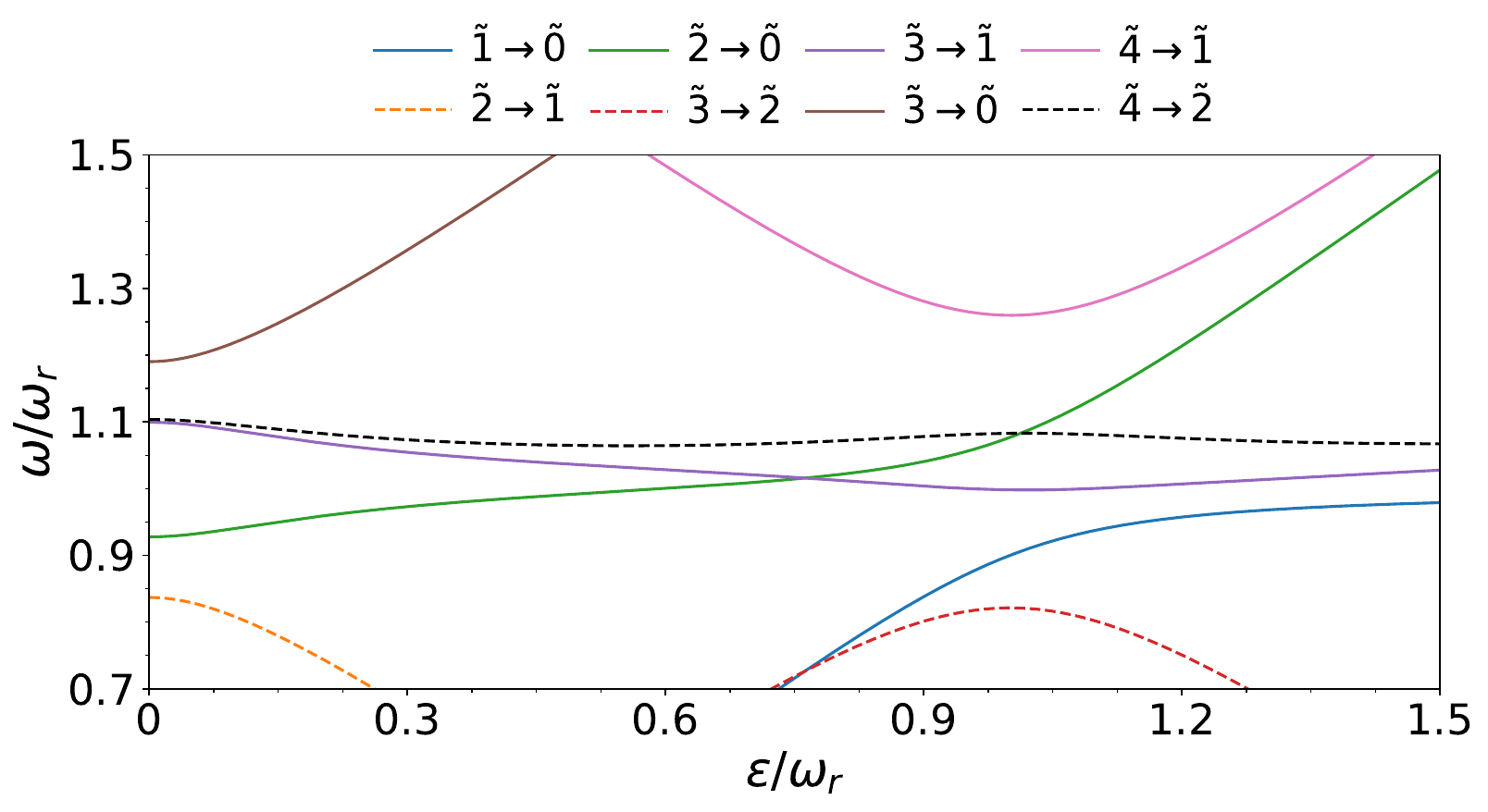}
    \caption{Energy transitions as a function of $\epsilon/\omega_r$ for $\eta = 1.01$ and $\Delta/\omega_r= 0.69$.} 
    \label{auto3}
\end{figure}

\begin{figure}[b]
    \centering
    \includegraphics[width = \linewidth]{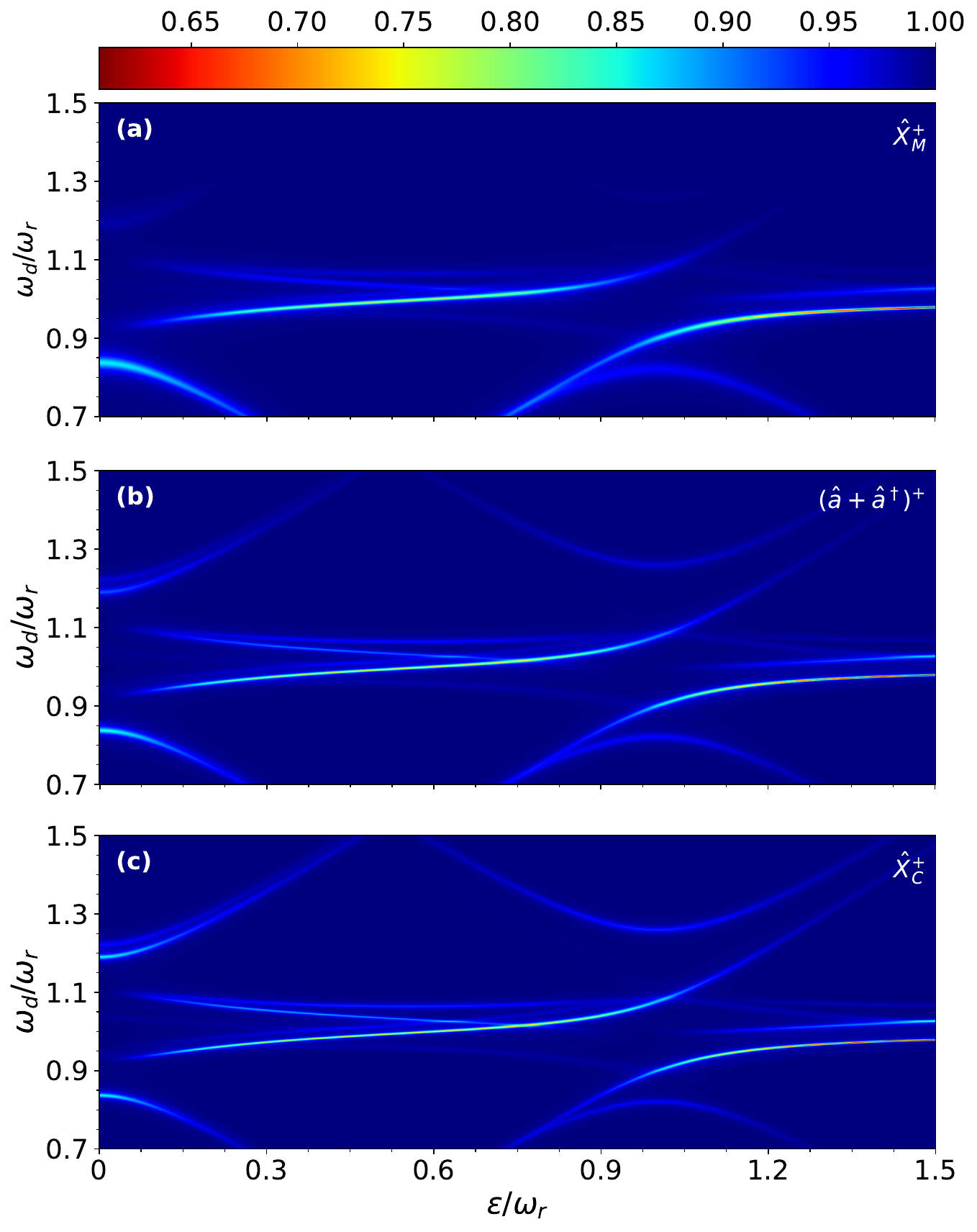}
    \caption{Coherent spectra obtained for $\eta = 1.01$, $\Delta/\omega_r= 0.69$. 
    (a) Reflectivity spectra $S^M_{11}(\omega_d)$ for $\hat X^{+}_M$. (b) $S^M_{11}(\omega_d)$ is evaluated for $(\hat a+ \hat a^\dagger)^{+}$. Reflectively spectra $S^C_{11}(\omega_d)$ for $\hat X^{+}_C$. The color map is set so that the maximum value is chosen to 1, and the lowest value as the absolute minimum among all spectra.}
    \label{rfs}
\end{figure}

In \figref{auto3}, we plot the transition energies as a function of the normalized flux offset $\epsilon/\omega_r$. They determine the main spectral lines in the considered spectral window (see \figref{rfs}). As expected, the transition energies in \figref{auto3} correspond to those observed in Refs.\,\cite{Yoshihara2017,Phys.Rev.Semba} (here, the considered spectral range is larger) and they will not be further discussed here.
Throughout this section, we use the following damping rates: $\gamma_m/\omega_r = \gamma_c/\omega_r= 10^{-3}$ , $\gamma_q/\omega_r = 0.005$. The effective temperatures are all put to $0.55$.
In order to also excite higher energy levels, $\abs{b_{\rm in}}$ is set to $0.03$.
All the coherent spectra have been obtained using the generalized Liouvillian described in \appref{appB}.

Figure~\ref{rfs}(a) shows the reflectivity  for the mutual inductive coupling with the TL ($S^M_{11}$) as function of $\omega_d/\omega_r$ and $\epsilon/\omega_r$. 
The transitions ($\tilde 1$, $\tilde 0$), ($\tilde 2$, $\tilde 0$), and ($\tilde 2$, $\tilde 1$) are those with the lowest reflectivity (the highest reduction of the input signal).

In References\,\cite{Phys.Rev.Semba,Yoshihara2017}, the experiment was conducted through a mutual inductive coupling between an LC oscillator and a coplanar TL, and their estimations on the spectra were made with $(\hat a+ \hat a^\dagger)$, so we also present this case in \figref{rfs}(b). Moreover, in \figref{rfs}(c), we display the reflectivity spectra ($S^C_{11}$) for the system capacitively coupled to the TL. The main differences between the three panels in \figref{rfs} can be associated to the ($\tilde 2$, $\tilde 0$), ($\tilde 3$, $\tilde 0$), and ($\tilde 4$, $\tilde 1$) spectral lines.

The transition ($\tilde 2$, $\tilde 0$) is visible up to $\epsilon/\omega_r = 1.5$ in both $S_{11}^M$, computed with $(\hat a + \hat a^\dagger)$, and $S_{11}^C$ spectra, as shown in \figref{rfs}(b) and (c). However, this spectral line quenches for $\epsilon/\omega_r \geq 1.2$ in the $S_{11}^M$ reflectivity spectra using $\hat X_M$ [see \figref{rfs}(a)].

The ($\tilde 3$, $\tilde 0$) spectral line displays a greater reduction of the input signal (i.e., lower reflectivity) in \figref{rfs}(c) compared to \figref{rfs}(b). Additionally, \figref{rfs}(a) highlights the highest reflectivity for this transition among the analyzed cases, with quenching observed for $\epsilon/\omega \geq 0.3$.

\begin{figure}[t]
    \centering
    \includegraphics[width = \linewidth]{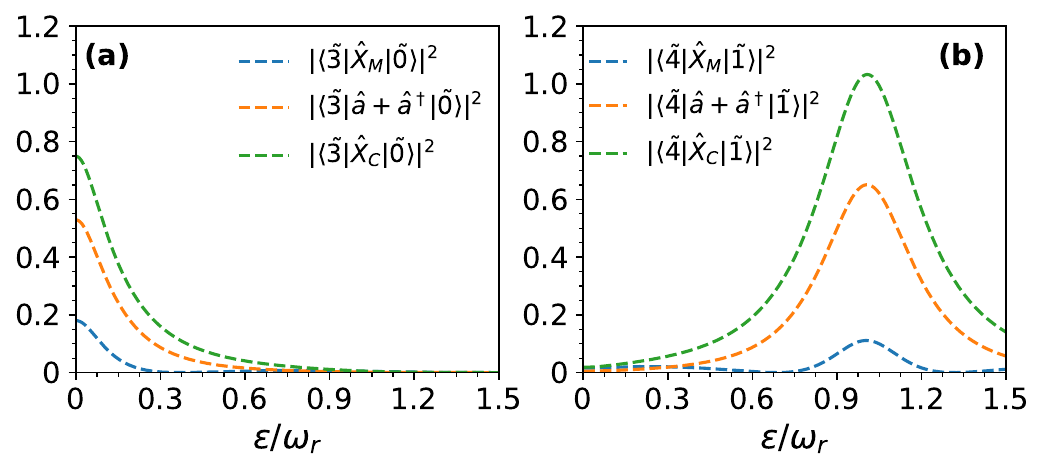}
    \caption{Squared modulus of three transition matrix elements as a function of $\epsilon/\omega_r$, with $\eta=1.01$ and $\Delta/\omega_r=0.69$. Panel (a) shows $\abs{\bra{\tilde 3} \hat{X}_{M(C)}\ket{\tilde 0}}^2$
    and $\abs{\bra{\tilde 3}(\hat a+ \hat a ^\dagger)\ket{\tilde 0}}^2$, while panel (b) reports $\abs{\bra{\tilde 4}\hat{X}_{M(C)}\ket{\tilde 1}}^2$ and $\abs{\bra{\tilde 4}(\hat a+ \hat a ^\dagger)\ket{\tilde 1}}^2$. }
    \label{mec}
\end{figure}

Finally, the ($\tilde 4$, $\tilde 1$) transition is not visible in \figref{rfs}(a), while it is present in both \figrefs{rfs}(b) and (c). All the previous observations are supported by the matrix elements shown in \figref{mec}.
These effects are not reported in the the Supplementary Materials of Ref.\,\cite{Yoshihara2017}, as this frequency region was not investigated there.

\section{Conclusions}
\label{sec Conclusion}

We presented a general framework for calculating spectra in circuit QED systems, which is also suitable for those operating in the USC and DSC regimes.
We applied it to calculate emission spectra in a system constituted by a flux qubit coupled to an LC electromagnetic resonator, under incoherent (thermal-like) excitation of the qubit. We find that, even at zero flux offset, when the energy levels correspond to those of the quantum Rabi model for dipolar atomic emitters (cavity QED), the circuit QED spectra may significantly differ from the corresponding cavity QED ones. In particular, we show that the circuit QED spectra can depend on how the resonator is coupled to the output port used for detection. In the case of a resonator capacitively coupled to the output port, cavity QED spectra are recovered in the absence of qubit flux offset.

We also computed reflectivity spectra under coherent drive. In this case as well, the spectra can depend on how the system is coupled to the input-output TL.

These differences between circuit and cavity QED spectra, even at zero flux offset, originate from different light-{\em matter} coupling Hamiltonian (or Lagrangian) term. In cavity-QED, the interaction Lagrangian term is of the form coordinate-momentum (field coordinate-matter momentum in the Coulomb gauge and conversely in the dipole gauge), see, e.g, Ref.\,\cite{PureDeph-Alberto}. In contrast, in the circuit QED systems here considered, the corresponding interaction term is of the form coordinate-coordinate (or momentum-momentum in the so-called charge gauge \cite{Manucharyan_2017}).

These results indicate that an accurate description of both incoherent and coherent spectra in the USC and DSC regimes requires taking into account the proper system variable which is probed. This framework can also be applied to study quantum nonlinear optics processes in ultrastrong circuit QED \cite{wang2024strong,Wang2023,tomonaga2024photonsimultaneouslyexcitesatoms}. Very recently, 
 it has been shown that spontaneous Raman scattering of incident radiation can be observed in USC cavity-QED systems without external enhancement or coupling to any vibrational degree of freedom \cite{Raman-Alberto}. Raman scattering processes can be evidenced as resonances in the emission spectrum, which become clearly visible as the cavity-QED system approaches the USC regime. It would be interesting to calculate the corresponding Raman spectra in circuit QED, which is an ideal platform to observe these effects.

\section{\label{sec: Acknowledgments}Acknowledgments}

We acknowledge the Army Research Office (ARO) (Grant No. W911NF-19-1-0065). We also acknowledge discussions with Fabio Mauceri.

\begin{appendices}


\section*{Appendices}
\addcontentsline{toc}{section}{Appendices}
\renewcommand{\thesection}{A.\arabic{section}}
\renewcommand{\thesubsection}{\alph{subsection}}
\section{Circuits quantization}\label{appA}

In this Appendix, we present the Lagrangian and the Hamiltonian of the circuital system considered throughout this work (see \figref{circuit}). Such circuit is subsequently coupled to TLs operating as input-output ports.

We consider a single flux qubit whose 
Lagrangian $\mathcal{L}_{q}$ is given by \cite{10.1063/1.2155757} 
$$ \mathcal{L}_{q}= 1/2 \; C_{j} \dot{\Phi}^2_{q} -\Phi^2_q/2\Tilde{L} + E_j \cos\bigr[2\pi/\Phi_0 \; (\Phi_q- \Phi_ {\rm ext})\bigl]\,,$$ where $\Tilde{L} \simeq (L\,L_C)/(L+L_C) $ is the equivalent inductance which does not take into account the contribute of the loop inductance $L_j$.

Considering that $C_K \ll \{C_T, C\}$, the total Lagrangian (consisting of the circuit QED system coupled capacitively to the input-output TL) is given by:
\begin{equation}\label{LagragT}
\begin{split}
\mathcal{L}_{C} &= \mathcal{L}_q + \frac{1}{L} \Phi \Phi_q + \frac{1}{2}\,   \dot{\underline{{\Phi}}}^T \, \underline{\underline{C}}\, \dot{\underline{{\Phi}}}-\frac{\Phi^2}{2L}- \frac{(\Phi_1-\Phi_0)^2}{2L_T }+\\
&+\sum_{i=1}^{\infty}\frac{C_T}{2}\dot{\Phi_i}^2 - \sum_{i=1}^{\infty}\frac{(\Phi_{i+1}-\Phi_i)^2}{2L_T }\,,
\end{split}
\end{equation}
where 
\begin{equation}
\dot{\underline{{\Phi}}}
=
\begin{pmatrix}
   \dot{\Phi}    \\
   \dot{\Phi}_0    
\end{pmatrix}\,, \; \;
\underline{\underline{C}} = 
   \begin{pmatrix}
  C+C_K  & -C_K\\
     -C_K &  C_T+C_K
    \end{pmatrix}\,. 
\end{equation}

From \eqref{LagragT}, it is possible to obtain the total Hamiltonian:

\begin{equation}\label{eq39}
\begin{split}
\mathcal{H}_{C} &=\mathcal{{H}}_{\rm 0} +\mathcal{{H}}_{\rm tl} + \frac{C_K}{C C_T} Q\,Q_0\,
\end{split}
\end{equation}
where
$$\mathcal{{H}}_{\rm 0}=\mathcal{H}_q + \mathcal{{H}}_{\rm LC} +\mathcal{{H}}_{g}$$
is the flux gauge Hamiltonian, including respectively the qubit Hamiltonian \cite{Yoshihara2022}
\begin{equation}\label{hamqubit}
\mathcal{{H}}_q =\frac{{Q}_q^2}{2C_j}+\frac{{\Phi}^2_q}{2 \Tilde{L}} - E_j \cos\left[{ \frac{2\, \pi}{\Phi_0}}({\Phi}_q-\Phi_{\rm ext})\right]\,,
\end{equation}
the LC oscillator term
\begin{equation}
\mathcal{{H}}_{LC} = \frac{{Q}^2}{2C}+\frac{{\Phi}^2}{2L}\,,
\end{equation}
 and the coordinate-coordinate (galvanic) interaction term
 $$\mathcal{{H}}_{g}=- \frac{1}{L} \Phi \Phi_q\,.$$
Moreover, the Hamiltonian of the TL is given by
\begin{equation*}
\begin{split}
\mathcal{{H}}_{\rm tl} &= \frac{1}{2 C_T} Q^2_{0}+
\frac{(\Phi_1-\Phi_0)^2}{2L_T }+\sum_{i=1}^{\infty}\frac{Q^2_i}{2C_T}\\
&+\sum_{i=1}^{\infty}\frac{(\Phi_{i+1}-\Phi_i)^2}{2L_T } \, .
\end{split}
\end{equation*}

We take the continuum limit of the TL and, as a consequence, we need to introduce the continuous quantities $\phi(x)$ for the flux field and  $\rho(x) = Q_n/\Delta x$  for a charge density field
\cite{RevModPhys.93.025005}. The canonical quantization can be performed by promoting the canonical variables to operators, that is $\comm{\hat{\Phi}}{\hat{Q}} = i $, $\comm{\hat{\Phi}_q}{\hat{Q}_q} = i $ and $\comm{\hat{\phi}(x)}{\hat \rho(x')} = i  \, \delta(x-x')$.

We now consider a semi-infinite TL, which acts as an input-output port.
In this particular case, we can expand the charge field operator as (see, e.g., Ref.\, \cite{RevModPhys.91.025005})
\begin{equation}\label{chargefield}
\hat{\rho}(x) = i \int_{0}^{\infty} d\omega \sqrt{\frac{ \omega \, c_T
}{\pi v_0 }}\, \cos{\biggl(\frac{\omega x}{v_0}\biggr)} (\hat b^{\dagger}_{\omega}- \hat b_{\omega})\,.
\end{equation}
In the LC circuit, the charge operator can be expressed as
\begin{equation}\label{chargeop}
\hat Q = i Q_{\rm zpf}\, (\hat a^\dagger-\hat a)\,,
\end{equation}
where  $Q_{\rm zpf} = \sqrt{(C \omega_r)/2}$. Thus, by substituting \eqref{chargeop} and \eqref{chargefield} in the last term of \eqref{eq39}, we recover \eqref{capfull} and \eqref{vcap} in the continuum limit of the TL.



Analogously to the capacitive case, we can obtain the full Hamiltonian for a mutual inductive coupling between the circuit and the TL [see \figref{circuit}(a)].
The total system Hamiltonian (consisting of the circuit QED system coupled inductively to the input-output TL) reads
\begin{equation}\label{incl}
\begin{split}
\mathcal{H}_ {M} &= \mathcal{H}_q + \frac{Q^2}{2\, C} + \frac{1}{2 \, \tilde{L}}\, \Phi^2 + \frac{1}{2\,L} (\Phi- \Phi_q)^2 + \frac{1}{2 \, C_T}\times \\ 
&\times \sum_{j =1} Q^2_j + \frac{1}{2\tilde L_T} (\Phi_{n+1}-\Phi_{n})^2 + \frac{1}{2\, L_T}\\ 
& \times \sum_{j \not = n} (\Phi_{j+1}-\Phi_{j})^2+\frac{\Phi-\Phi_q}{\tilde{M}}\, (\Phi_{n+1}-\Phi_{n})\,.
\end{split}
\end{equation}
Note that from last term in the final row of \eqref{incl},
we can argue how the coupling with the TL is obtained through an effective mutual inductance $\tilde M = (L\, L_T-M^2)/M$ ( Ref.\,\cite{10.1063/1.5089550}).
Furthermore,  note that we have introduced in \eqref{incl} the rescaled inductances $ \tilde L=(L\, L_T-M^2)/L_T$ and $ \tilde L_T= (L\, L_T-M^2)/L$. 


Considering the infinite extension of the TL, performing the continuum limit and the quantization procedure, the flux field $\hat{\phi}(x)$ operator can be expanded as \cite{RevModPhys.93.025005}
\begin{equation}
\hat{\phi}(x) = \int_{0}^{+\infty} d\omega\, \frac{\Lambda}{\sqrt{\omega}}\big(\hat{b}_\omega \, e^{i k x} +\hat{b}^\dagger_\omega \, e^{-i k x} \big)\,, 
\end{equation}
where $k= \omega/v_0$ is the wavenumber.
The interaction Hamiltonian between the system and the TL, $\hat{\mathcal V}_{M}$, can be expressed as
\begin{equation}
\hat{\mathcal V}_{M}=  \alpha  \hat \Phi_L\, \int dx\, \delta(x) \frac{\partial \hat{\phi}(x)}{\partial x}\,,
\end{equation}
where $$\hat \Phi_L= \hat \Phi - \hat \Phi_q = \Phi_{\rm zpf}\, [\hat a + \hat{a}^\dagger-2\eta\, (\cos{\theta}\, \hat{\sigma}_z -\sin{\theta} \,\hat{\sigma}_x)]\,.$$
In the end, we find:
\begin{equation}\label{inductinter}
\hat{\mathcal V}_{M}= i\, \frac{1}{\Phi_{ \rm zpf}} \, \hat{\Phi}_L \int_{0}^{+\infty} d\omega\, \alpha \Phi_{ \rm zpf}\, \frac{\Lambda}{v_0}\, \sqrt{\omega} \, [\hat{b}_\omega-\hat{b}^\dagger_\omega ]\,,
\end{equation}
so  \eqref{qmi} and \eqref{vind} are demonstrated.

\section{Input-Output theory}\label{appIO}
In this Appendix, we derive the input-output relations in order to establish which system operator is related to the voltage measured for both coupling schemes.

In the case of the infinite TL connected to the qubit-LC system through a mutual inductance, we find that the continuum limit of the Hamiltonian of \eqref{incl} reads as

\begin{equation}
\begin{split}
\hat {\mathcal{H}}_M = \hat {\mathcal{H}}_{0} &+ \int dx \left( \frac{\hat{\rho}^2_c(x)}{2 c_T}+ \frac{(\partial_x \hat{\phi})^2}{2 l_T}\right)\\ &+\frac{M}{L l_T} \hat \Phi_L \int \delta(x)\partial_x \hat{\phi}\,.
\end{split}
\end{equation}
Therefore, the resulting Heisenberg equations of motion are
\begin{equation}\label{fpunto}
\dot{\hat{\phi}}(x) =  \frac{\hat{\rho}(x)}{ c_T}\,
\end{equation}
and
\begin{equation}
\begin{split}
\dot{\hat{\rho}}(x)&=\int dx' \frac{i}{l_T } \partial_{x'} \hat{\phi}  \comm{\partial_{x'} \hat{\phi}}{\hat{\rho}(x)} + \frac{i M}{L l_T} \delta(x') \comm{\partial_{x'} \hat{\phi}}{\hat{\rho}(x)} \\
&= -\int dx' \frac{1}{l_T } \left(\partial_{x'} \hat{\phi}\right) \, \partial_{x'} \delta(x'-x)+ \\
&\; \; \; \,-\int dx'\frac{M}{L\,l_T } \delta(x') \partial_{x'} \delta(x'-x) \hat{\Phi}_L = \\
&= \frac{1}{l_T} (\partial^2_x \hat{\phi})+ \frac{M}{L l_T} \delta'(x) \hat \Phi_L\,.
\end{split}
\end{equation}
Substituting this last equation in \eqref{fpunto}, we obtain 
\begin{equation}
\ddot{\hat{\phi}}(x)= \frac{\dot{\hat{\rho}}(x)}{ c_T} = v^2_0 (\partial^2_x \hat{\Phi})+ \frac{M}{c_T \, L l_T} \delta'(x) \hat{\Phi}_L\,,
\end{equation}
corresponding in the Fourier space to
\begin{equation}
\biggl(k^2 +  \partial^2_x\biggr) \hat \phi(x,\omega) = - \frac{M}{v_0^2\, c_T L l_T} \hat{\Phi}_L(\omega) \delta'(x)\,.
\end{equation}
The solution to this differential equation can be expressed by means of the Green's functions \cite{economou2006green}:
\begin{equation}
\hat{\phi}^{\pm}_{M_{\rm out}}(x,\omega) = \hat{\phi}^{\pm}_{\rm in} (x,\omega) - \biggl( \frac{ \, M}{2 v^2_0 l_T L c_T} e^{\pm
 i\, k \abs{x}}\hat{\Phi}^{\pm}_L(\omega)\biggr)\,.
\end{equation}
Thus, since $\hat{V}^{\pm}_{M_{\rm out}}= \dot{\hat{\phi}}^{\pm}_{M_{\rm out}} $, we obtain the output voltage as 
\begin{equation} 
\hat{V}^{\pm}_{M_{\rm out}}(x, \omega) = \hat{V}^{\pm}_{\rm in} (x,\omega)\pm i \omega\,\frac{ M}{2  L }  e^{\pm
 i\, k \abs{x}}  \, \hat{\Phi}^{\pm}_L(\omega)\,.
\end{equation}
The last equation can be written in the time domain as 
\begin{equation}
\hat{V}^{\pm}_{M_{\rm out}}(x, t) =  \hat{V}^{\pm}_{\rm in} (x,t)- \,\frac{ M}{2  L }   \, \dot{\hat{\Phi}}^{\pm}_L(x, t)\,,
\end{equation}
where 
\begin{equation}
\hat{\Phi}^{\pm}_L(x, t) = \int_0^{\infty} d \omega \, e^{
\pm i\, (\omega/v_0)\abs{x}}  \, \hat{\Phi}^{\pm}_L(\omega) e^{\mp i \omega t}\,.
\end{equation}

For the capacitive coupling with the semi-infinite TL, following the quantum network theory and imposing the boundary condition at $x  = 0$  \cite{RevModPhys.93.025005,Quantumnet,gardiner2004quantum}, 
the resulting  output voltage can be written as
\begin{equation}
\hat{V}^{\pm}_{C_{\rm out}}(t - x/v_0) = \hat{V}^{\pm}_{in}(t+x/v_0) + Z_0 \frac{C_K}{C} \dot{\hat{Q}}^{\pm}(t - x/v_0)\,.
\end{equation}

\section{Generalized Liouvillian }\label{appB}
In this appendix, we explicitly present the full form of the generalized Liouvillian used in \eqref{lou}.
Following the derivation in Ref.\,\cite{PhysRevA.98.053834}, we obtain
\begin{align} \label{gmemi}
\mathcal{L}_{ g} \, \hat \rho &= \frac{1}{2}\sum_{\substack{i = (r,q) \\ (\omega, \omega')>0}} \bigg\{ \frac{\gamma_i \,\omega'}{\omega_i}\, n_{\rm th}(\omega',T_{i}) \bigl[ \hat A_i^{(-)} (\omega') \hat \rho (t) \hat A_i^{(+)} (\omega)\nonumber\\
&- \hat A_i^{(+)} (\omega) \hat A_i^{(-)} (\omega') \hat \rho (t) \bigr]   +  \frac{\gamma_i \,\omega}{\omega_i} \, n_{\rm th}(\omega,T_{i})  \nonumber \\
&\times \bigl[ \hat A_i^{(-)} (\omega') \hat \rho (t) \hat A_i^{(+)} (\omega) - \hat \rho (t)\hat A_i^{(+)} (\omega) \hat A_i^{(-)} (\omega')\bigr] \nonumber\\
& + \frac{\gamma_i \,\omega}{\omega_i} [n_{\rm th}(\omega,T_{i}) + 1] \bigl[ \hat A_i^{(+)} (\omega) \hat \rho(t) \hat A_i^{(-)} (\omega')\nonumber \\
&- \hat A_i^{(-)} (\omega') \hat A_i^{(+)}(\omega) \hat \rho (t)\bigr] +  \frac{\gamma_i \,\omega'}{\omega_i}\, [n_{\rm th}(\omega',T_{i}) + 1]\nonumber\\ 
& \times\bigl[ \hat A_i^{(+)} (\omega) \hat \rho (t) \hat A_i^{(-)} (\omega') - \hat \rho (t) \hat A_i^{(-)} (\omega') \hat A_i^{(+)} (\omega) \bigr] \nonumber \\
& + \frac{\gamma_q}{\Delta}\,T_q \, \mathcal{D}[\hat{\tilde{\sigma}}_x^{(0)}] \hat \rho (t)+ \frac{\gamma_q}{\Delta}\,(T_q+1)\,\mathcal{D}[\hat{\tilde{\sigma}}_x^{(0)}]\hat \rho (t)\,,
\end{align}  
where $n_{\rm th}(\omega,T_i)= (e^{{\omega}/{ T_i}}-1)^{-1}$ describes the mean number of thermal photons and $\omega$ (or $\omega'$) is used for the transition frequencies of the light-matter system.
$\hat A^{+(-)}_{i}$ denotes the positive (or negative) frequency component of the corresponding operator written in the diagonal basis of $\mathcal{\hat{H}}_0$ [see \eqref{qb-LC}].
Using the following relation for the diagonal terms of a generic system operator
\begin{equation}
\hat{S}^{(0)}=\sum_{i}  {S}_{ii}\,\ketbra{i}{i}\,,  
\end{equation}  
the last row of \eqref{gmemi} can be shown to describe the pure dephasing effects associated to the zero-frequency component of $\hat{\tilde{\sigma}}_x$, where 
\begin{equation}
\mathcal{D}[\hat {\tilde{\sigma}}_x^{(0)}] \hat \rho = \hat{\tilde{\sigma}}_x^{(0)}  \hat \rho \, \hat{\tilde{\sigma}}_x^{(0)}  - \frac{1}{2} \hat{\tilde{\sigma}}_x^{(0)}  \hat{\tilde{\sigma}}_x^{(0)} \,\hat \rho -\frac{1}{2} \,\hat \rho \, \hat{\tilde{\sigma}}_x^{(0)}  \, \hat{\tilde{\sigma}}_x^{(0)} 
\end{equation}
is the Lindblad dissipator. If the symmetry is not broken ($\epsilon=0$), this term goes to zero.
In \eqref{gmemi}, for $i = r $, $\omega_i$ corresponds to the LC resonance frequency $\omega_r$, whereas, for $i=q$, $\omega_i$ is associated to $\Delta$.
Moreover, the operator $\hat A_{r}$ corresponds to $\hat X_C$,  $\hat X_M$ or $\hat{X}_D$ in the case of the capacitive coupling, inductive coupling or cavity QED, respectively.
Concerning the qubit internal losses, $\hat A_q = \hat{\tilde{\sigma}}_x $ for both circuital schemes, whereas $\hat A_q =\hat \sigma_x$ when we consider the cavity QED dissipation (see \secref{i/o_rel} and \secref{mees}).

In the interaction picture, the terms in the first six rows of \eqref{gmemi} oscillate at frequencies $\pm(\omega-\omega')$. As shown in Ref.\,\cite{PhysRevA.98.053834}, if such difference is much greater than the damping rates, these highly oscillating terms can be neglected. Their numerical computation can give rise to singularities in the logarithmic spectra, which can be prevented by the introduction of a Gaussian numerical filter:
\begin{equation}
F(\omega, \omega') = e^{-\frac{\abs{\omega-\omega'}^2}{2 b^2}}\, .
 \end{equation}
The generalized master equation can be reconnected to the dressed master equation found in Ref.\,\cite{PhysRevA.84.043832} for very low values of $b$. The GME works adequately for quantum optical systems displaying hybrid harmonic-anharmonic energy specta and for whatever coupling strength regime, even if the Lindblad form is lost at $T \neq 0$.

In \secref{cohspect}, we derived the reflectivity using a master equation approach. In contrast to incoherent processes, the Liouville equation can now be expressed as
\begin{equation}\label{lou2}
\dot{\hat \rho} = - i \comm{\mathcal{\hat H}_0}{\hat \rho}+(\mathcal{L}_{g}+ \mathcal{L}_+ e^{i \omega_d t} + \mathcal{L}_- e^{-i \omega_d t}  )\, \hat \rho\,,
\end{equation}
where $\mathcal{L}_{g}$ is related to the dissipation channels of the system and it has the form of \eqref{gmemi}.
The terms $\mathcal{L}_{\pm}$ in \eqref{lou2} originates from driving coherently the system. In particular, 
\bea 
\mathcal{L}_{\pm} \hat \rho = \pm i \abs{b_{\rm in}}e^{\pm i \varphi} \sqrt{\frac{\gamma_{c}\,\omega_d}{\omega_r}} \comm{\hat X_C}{\hat \rho} \label{h+}\,,   \\
\mathcal{L}_{\pm} \hat \rho = \mp i \abs{b_{\rm in}}e^{\pm i \varphi} \sqrt{\frac{\gamma_{m}\, \omega_d}{\omega_r}} \comm{\hat X_M}{\hat \rho} \label{h-}\,,
\eea
where \eqref{h+} holds for the capacitive coupling and \eqref{h-} refers to the mutual inductive coupling case. The variable $\varphi$ is the phase of the driving tone. The opposite sign in these equations is due to the nature of the interaction Hamiltonian [see \eqref{vind} and \eqref{vcap}].

Substituting the Fourier expansion of the steady state density matrix, $\hat{\rho}_{ss}(t)= \sum_{k = -\infty}^{+\infty}\hat \rho^{k}\, e^{i\,k\omega_d\,t}$, in \eqref{lou2}, we obtain the following recursive equation \cite{wang2024strong}:
\begin{equation}
(\mathcal{L}_g- i k \, \omega_d)\hat \rho ^k+ \mathcal{L}_+ \hat{\rho}^{k-1}+ \mathcal{L}_- \hat \rho^{k+1} = 0\,.
\end{equation}
The various components of $\rho ^k$ can be numerically computed by solving the last relation, and truncating the Fourier series to the desired order \cite{wang2024strong}.

\section{Further Spectra}\label{appspec}
\begin{figure}[htb]
    \centering
    \includegraphics[width = 0.5\textwidth]{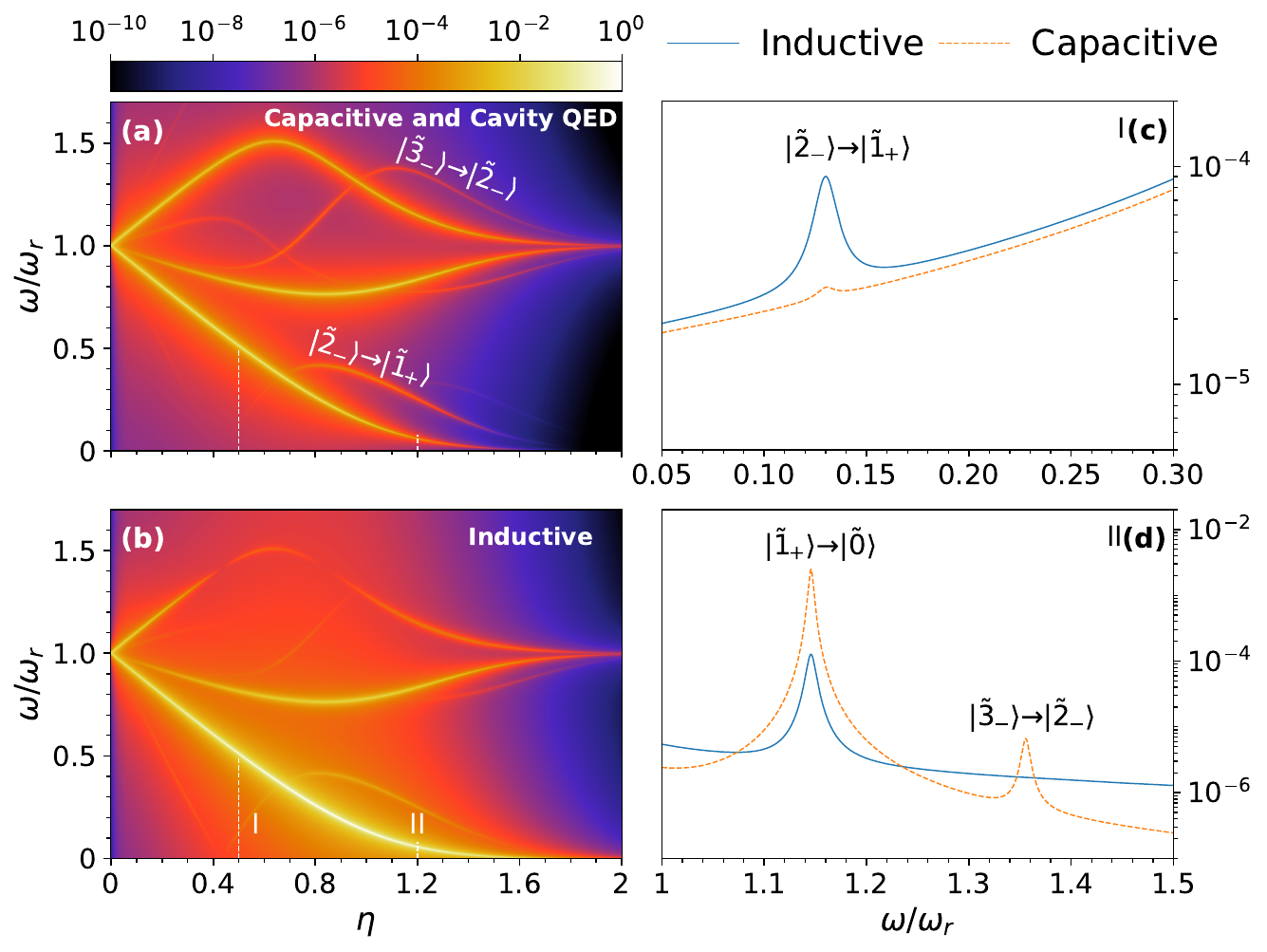}
    \caption{Logarithmic emission spectra obtained for $T_q/\omega_r= 0.18$, $\Delta/\omega_r = 1$, and $\epsilon = 0$. (a) Logarithmic spectra $S_C(\omega)$ as a function of $\eta$, for the system capacitively coupled to the TL (also valid for the cavity QED case). (b) Logarithmic spectra for the mutual inductive coupling,  $S_M(\omega)$, with the TL. Spectra (a) and (b) are normalized with respect to the absolute maximum value. (c) Logarithmic spectrum computed for $\eta=0.5$, the capacitive coupling emission spectrum is enhanced by a factor of ten. (d) In this case, the logarithmic spectrum is evaluated at $\eta= 1.2$. Spectra (c) and (d) are normalized with respect to the maximum value of spectra (a) and (b).}
    \label{sp1a}
\end{figure}

In this Appendix, we extend the treatment discussed in \secref{emissionspectra}. Specifically, we will considered higher excitation strength ($T_q$) for the parity symmetry emission spectra. Subsequently, we will calculate emission spectra varying $\epsilon$, with fixed $\Delta=1$ (as displayed in \secref{ediv0}), for higher couplings $\eta$ in the DSC regime. In the end, further incoherent spectra will be presented for different values of $\Delta/\omega_r$.

\subsection{Parity symmetry (zero flux-offset)}
In this section, we present emission spectra obtained using the same parameters considered in \secref{zeroep} with a different effective temperature, $T_q=0.18$ (see \figref{sp1a}).

In contrast to \figref{fig1}, the $(\tilde 1_+,\tilde 0)$ spectral line is also visible for coupling strengths ranging from 0.4 to 1 (see \figref{sp1a}) due to the higher qubit effective temperature $T_q$. However, when  $\dot{\hat X}_M$ matrix elements vanish for this particular transition [see \figref{m1}(b)], we observe a suppression of the mutual inductive power spectra.

\begin{figure}[t]
    \centering
    \includegraphics[width= 1 \linewidth]{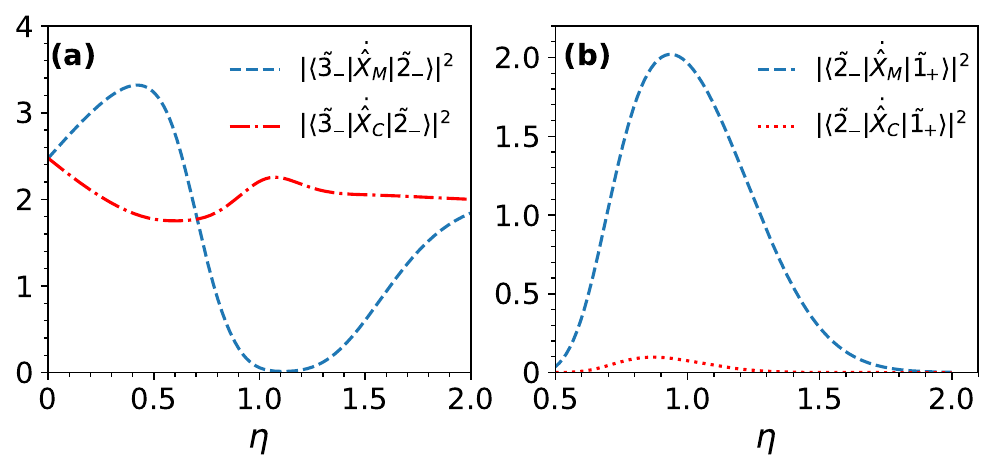}
    \caption{Modulus squared of two transitions matrix elements as a function of $\eta$, where $\epsilon=0$ and $\Delta/\omega_r=1$. (a) This panel shows $\abs{\bra{\tilde3_-}\dot{\hat{X}}_{M(C)}\ket{\tilde 2_-}}^2$, while panel (b) reports $\abs{\bra{\tilde2_-}\dot{\hat{X}}_{M(C)}\ket{\tilde 1_+}}^2$.} 
    \label{m1a}
\end{figure}

Two additional transitions have been highlighted: $(\tilde 2_-,\tilde 1_+)$ and $(\tilde 3_-,\tilde 2_-)$. When $\eta$ reaches 0.5, the former is noticeable principally for the mutual inductive spectrum. For $\eta = 1.2$, the transition $(\tilde 3_-,\tilde 2_-)$ is present only for the capacitive coupling spectrum [see \figrefs{sp1a}(c) and (d)]. The relative matrix elements of \figref{m1a} explain this behaviour.

 \subsection{Symmetry breaking (non-zero flux-offset)}

In this section, we present additional emission spectra as a function of the flux offset, considering different values of the normalized coupling strength. 

\begin{figure}[b]
    \centering
    \includegraphics[width = 0.5 \textwidth]{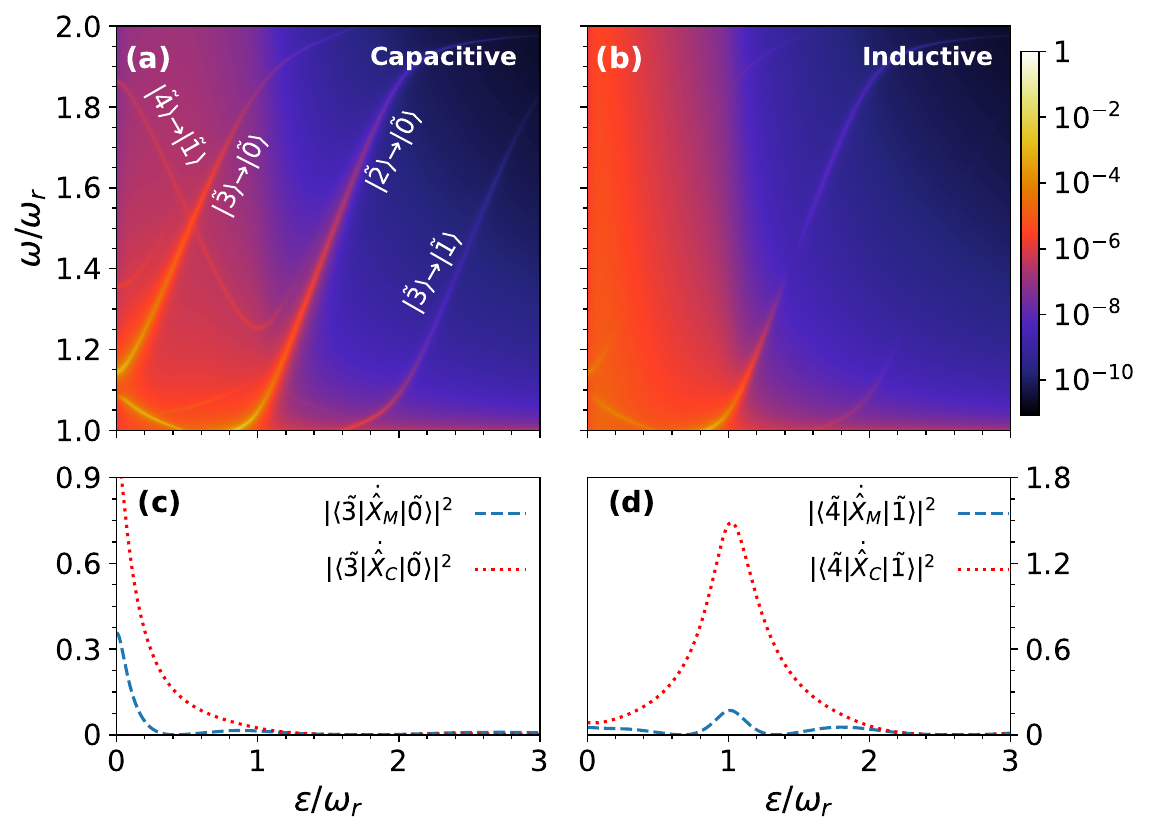}
    \caption{Logarithmic emission spectra obtained for $T_q/\omega_r= 0.15$, $\Delta/\omega_r = 1$, and $\eta = 1.2$. (a) Logarithmic spectra $S_C(\omega)$ as a function of $\epsilon/\omega_r$, for the system capacitively coupled to the TL. (b) Logarithmic spectra for the mutual inductive coupling,  $S_M(\omega)$, with the TL. Spectra (a) and (b) are normalized with respect to the absolute maximum value. The specific spectral lines, discussed in the text, are indicated in the panel (a). Panels (c) and (d) report two relevant transition matrix elements of $\dot{\hat X}_C$ and $\dot{\hat X}_M$ (the transitions displaying differences, see text).}
    \label{sp2a}
\end{figure}

Figure~\ref{sp2a} displays results obtained for $\eta =1.2$. The spectral lines corresponding to the transitions 
$(\tilde 2,\tilde 0)$ and $(\tilde 3,\tilde 1)$ exhibit similar features of \figref{2parte}. However, the transitions $(\tilde 3,\tilde 0)$ and $(\tilde 4,\tilde 1)$ are more visible for the capacitive coupling scheme [see \figrefs{sp2a}(a) and (b)]. Matrix elements for these transitions, which are shown in \figrefs{sp2a}(c) and (d), explain the origin of differences between  \figrefs{sp2a}(a) and (b).

\begin{figure}[ht]
    \centering
    \includegraphics[width = 0.5 \textwidth]{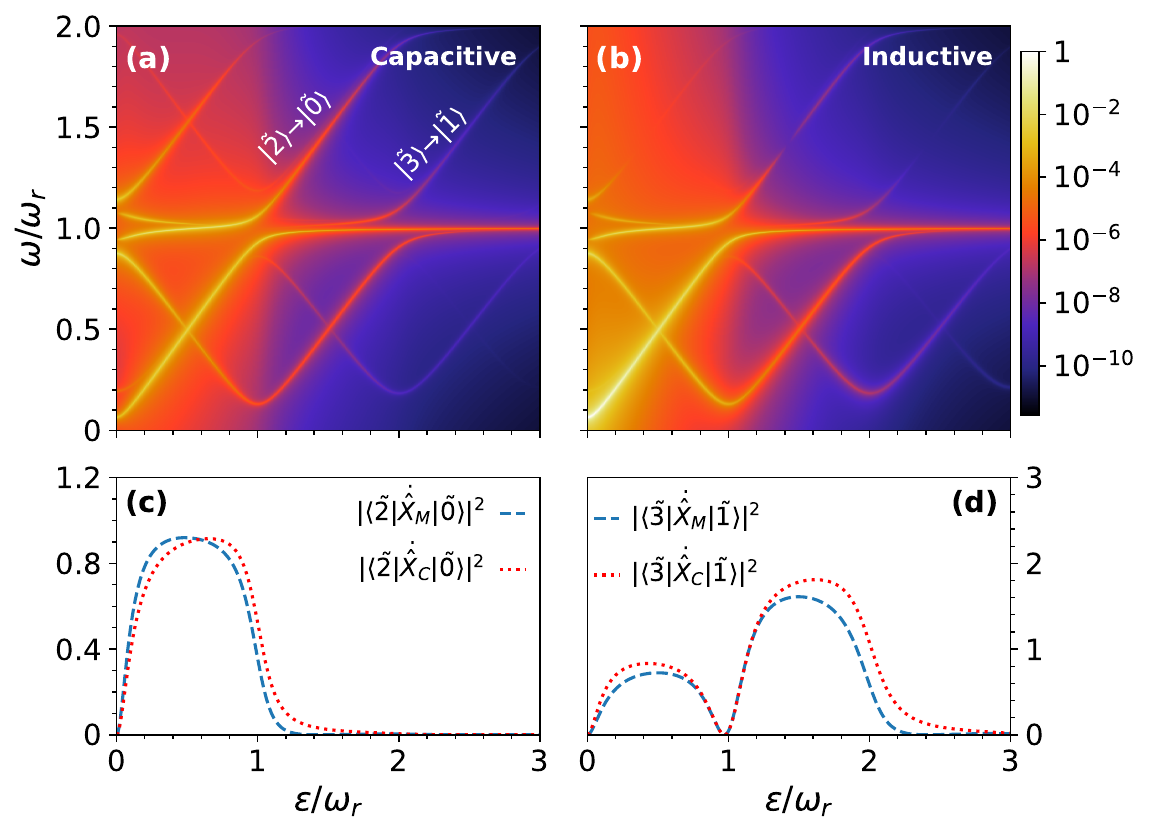}
    \caption{Logarithmic emission spectra obtained for $T_q/\omega_r= 0.15$, $\Delta/\omega_r = 0.5$, and $\eta = 1$. (a) Logarithmic spectra $S_C(\omega)$ as a function of $\epsilon/\omega_r$, for the system capacitively coupled to the TL. (b) Logarithmic spectra for the mutual inductive coupling,  $S_M(\omega)$, with the TL. Spectra (a) and (b) are normalized with respect to the absolute maximum value. The specific spectral lines, discussed in the text, are indicated in the panel (a). Panels (c) and (d) report two relevant transition matrix elements of $\dot{\hat X}_C$ and $\dot{\hat X}_M$ (the transitions displaying differences, see text).}
    \label{sp3a}
\end{figure}

\begin{figure}[H]
    \centering
    \includegraphics[width = 0.5 \textwidth]{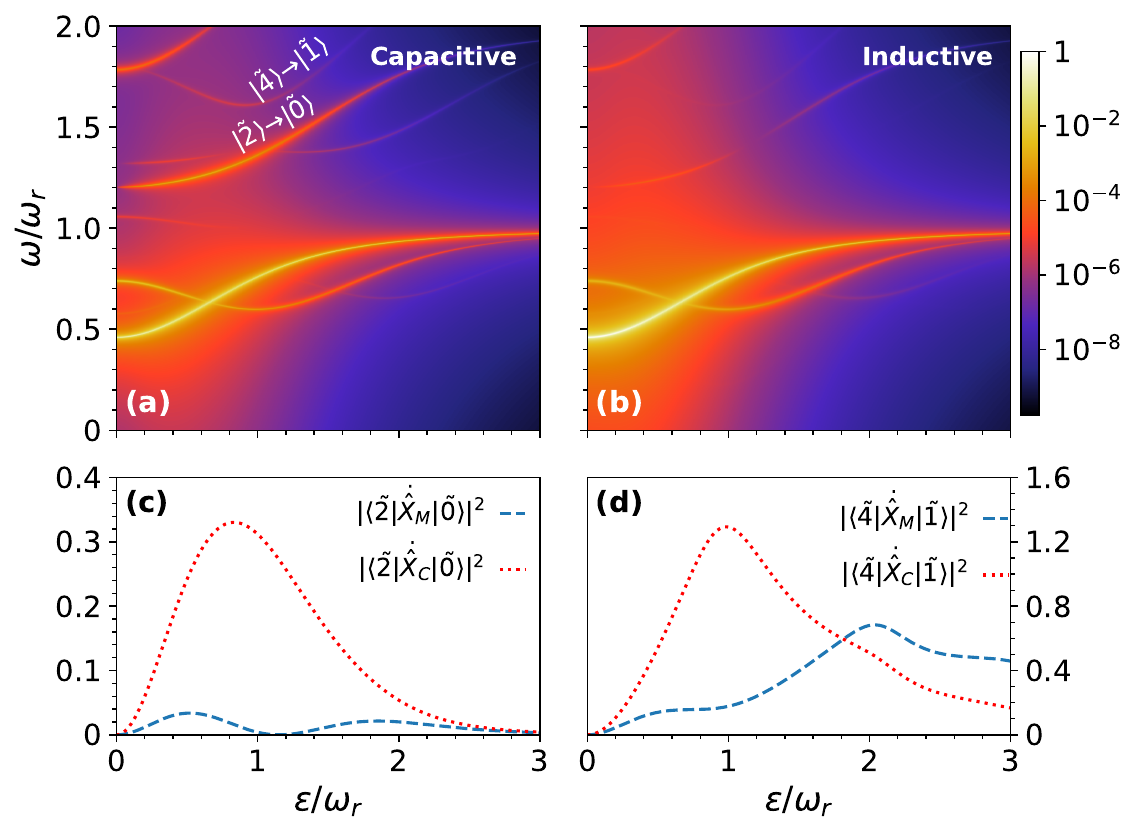}
    \caption{Logarithmic emission spectra obtained for $T_q/\omega_r= 0.15$, $\Delta/\omega_r = 1.5$, and $\eta = 0.7$. (a) Logarithmic spectra $S_C(\omega)$ as a function of $\epsilon/\omega_r$, for the system capacitively coupled to the TL. (b) Logarithmic spectra for the mutual inductive coupling,  $S_M(\omega)$, with the TL. Spectra (a) and (b) are normalized with respect to the absolute maximum value. The specific spectral lines, discussed in the text, are indicated in the panel (a). Panels (c) and (d) report two relevant transition matrix elements of $\dot{\hat X}_C$ and $\dot{\hat X}_M$ (the transitions displaying differences, see text).}
    \label{sp4a}
\end{figure}

\subsection{Varying  $\Delta$}

Throughout this work we have used $\Delta/\omega_r =1$ for all incoherent spectra.
Figure~\ref{sp3a} is obtained for $\Delta/\omega_r = 0.5$ and $\eta = 1$. In this case, there are still differences between the spectra in \figref{sp3a}(a) and (b) observed for particular spectral lines, which are reported in \figref{sp3a}(a), despite the very small differences in the matrix elements [see \figrefs{sp3a}(c) and (d)].

Emission spectra of \figrefs{sp4a}(a) and (b) have been obtained for $\Delta/\omega_r=1.5$ and $\eta= 0.7$. Also in this case, it is interesting to compare some differences in intensity of particular spectral lines with the corresponding matrix elements in \figrefs{sp4a}(c) and (d).

\end{appendices}


\newpage

\bibliography{bibliografia.bib}

\begin{thebibliography}{62}%
\makeatletter
\providecommand \@ifxundefined [1]{%
 \@ifx{#1\undefined}
}%
\providecommand \@ifnum [1]{%
 \ifnum #1\expandafter \@firstoftwo
 \else \expandafter \@secondoftwo
 \fi
}%
\providecommand \@ifx [1]{%
 \ifx #1\expandafter \@firstoftwo
 \else \expandafter \@secondoftwo
 \fi
}%
\providecommand \natexlab [1]{#1}%
\providecommand \enquote  [1]{``#1''}%
\providecommand \bibnamefont  [1]{#1}%
\providecommand \bibfnamefont [1]{#1}%
\providecommand \citenamefont [1]{#1}%
\providecommand \href@noop [0]{\@secondoftwo}%
\providecommand \href [0]{\begingroup \@sanitize@url \@href}%
\providecommand \@href[1]{\@@startlink{#1}\@@href}%
\providecommand \@@href[1]{\endgroup#1\@@endlink}%
\providecommand \@sanitize@url [0]{\catcode `\\12\catcode `\$12\catcode `\&12\catcode `\#12\catcode `\^12\catcode `\_12\catcode `\%12\relax}%
\providecommand \@@startlink[1]{}%
\providecommand \@@endlink[0]{}%
\providecommand \url  [0]{\begingroup\@sanitize@url \@url }%
\providecommand \@url [1]{\endgroup\@href {#1}{\urlprefix }}%
\providecommand \urlprefix  [0]{URL }%
\providecommand \Eprint [0]{\href }%
\providecommand \doibase [0]{http://dx.doi.org/}%
\providecommand \selectlanguage [0]{\@gobble}%
\providecommand \bibinfo  [0]{\@secondoftwo}%
\providecommand \bibfield  [0]{\@secondoftwo}%
\providecommand \translation [1]{[#1]}%
\providecommand \BibitemOpen [0]{}%
\providecommand \bibitemStop [0]{}%
\providecommand \bibitemNoStop [0]{.\EOS\space}%
\providecommand \EOS [0]{\spacefactor3000\relax}%
\providecommand \BibitemShut  [1]{\csname bibitem#1\endcsname}%
\let\auto@bib@innerbib\@empty
\bibitem [{\citenamefont {You}\ and\ \citenamefont {Nori}(2011)}]{Nori-atomic-physics}%
  \BibitemOpen
  \bibinfo {author} {J.~Q. You}\ and\ \bibinfo {author} {F.~Nori},\ \emph {\bibinfo {title} {Atomic physics and quantum optics using superconducting circuits}},\ \href {\doibase https://doi.org/10.1038/nature10122} {\bibfield  {journal} {\bibinfo  {journal} {Nature}\ }\textbf {\bibinfo {volume} {474}},\ \bibinfo {pages} {589} (\bibinfo {year} {2011})}\BibitemShut {NoStop}%
\bibitem [{\citenamefont {Dowling}\ and\ \citenamefont {Milburn}(2003)}]{Dowling2003}%
  \BibitemOpen
  \bibinfo {author} {J.~P. Dowling}\ and\ \bibinfo {author} {G.~J. Milburn},\ \emph {\bibinfo {title} {Quantum technology: the second quantum revolution}},\ \href {\doibase 10.1098/rsta.2003.1227} {\bibfield  {journal} {\bibinfo  {journal} {Philosophical Transactions of the Royal Society of London. Series A: Mathematical, Physical and Engineering Sciences}\ }\textbf {\bibinfo {volume} {361}},\ \bibinfo {pages} {1655} (\bibinfo {year} {2003})}\BibitemShut {NoStop}%
\bibitem [{\citenamefont {You}\ and\ \citenamefont {Nori}(2005)}]{10.1063/1.2155757}%
  \BibitemOpen
  \bibinfo {author} {J.~Q. You}\ and\ \bibinfo {author} {F.~Nori},\ \emph {\bibinfo {title} {{Superconducting Circuits and Quantum Information}}},\ \href {\doibase 10.1063/1.2155757} {\bibfield  {journal} {\bibinfo  {journal} {Physics Today}\ }\textbf {\bibinfo {volume} {58}},\ \bibinfo {pages} {42} (\bibinfo {year} {2005})}\BibitemShut {NoStop}%
\bibitem [{\citenamefont {Blais}\ \emph {et~al.}(2021)\citenamefont {Blais}, \citenamefont {Grimsmo}, \citenamefont {Girvin},\ and\ \citenamefont {Wallraff}}]{RevModPhys.93.025005}%
  \BibitemOpen
  \bibinfo {author} {A.~Blais}, \bibinfo {author} {A.~L. Grimsmo}, \bibinfo {author} {S.~M. Girvin},\ and\ \bibinfo {author} {A.~Wallraff},\ \emph {\bibinfo {title} {Circuit quantum electrodynamics}},\ \href {\doibase 10.1103/RevModPhys.93.025005} {\bibfield  {journal} {\bibinfo  {journal} {Rev. Mod. Phys.}\ }\textbf {\bibinfo {volume} {93}},\ \bibinfo {pages} {025005} (\bibinfo {year} {2021})}\BibitemShut {NoStop}%
\bibitem [{\citenamefont {Vool}\ and\ \citenamefont {Devoret}(2017)}]{Vool_2017}%
  \BibitemOpen
  \bibinfo {author} {U.~Vool}\ and\ \bibinfo {author} {M.~Devoret},\ \emph {\bibinfo {title} {Introduction to quantum electromagnetic circuits}},\ \href {\doibase 10.1002/cta.2359} {\bibfield  {journal} {\bibinfo  {journal} {International Journal of Circuit Theory and Applications}\ }\textbf {\bibinfo {volume} {45}},\ \bibinfo {pages} {897} (\bibinfo {year} {2017})}\BibitemShut {NoStop}%
\bibitem [{\citenamefont {Gu}\ \emph {et~al.}(2017)\citenamefont {Gu}, \citenamefont {Kockum}, \citenamefont {Miranowicz}, \citenamefont {Liu},\ and\ \citenamefont {Nori}}]{Gu2017}%
  \BibitemOpen
  \bibinfo {author} {X.~Gu}, \bibinfo {author} {A.~F. Kockum}, \bibinfo {author} {A.~Miranowicz}, \bibinfo {author} {Y.-X. Liu},\ and\ \bibinfo {author} {F.~Nori},\ \emph {\bibinfo {title} {Microwave photonics with superconducting quantum circuits}},\ \href {\doibase https://doi.org/10.1016/j.physrep.2017.10.002} {\bibfield  {journal} {\bibinfo  {journal} {Physics Reports}\ }\textbf {\bibinfo {volume} {718-719}},\ \bibinfo {pages} {1} (\bibinfo {year} {2017})}\BibitemShut {NoStop}%
\bibitem [{\citenamefont {Frunzio}\ \emph {et~al.}(2005)\citenamefont {Frunzio}, \citenamefont {Wallraff}, \citenamefont {Schuster}, \citenamefont {Majer},\ and\ \citenamefont {Schoelkopf}}]{Frunzio-characterizationofCQED}%
  \BibitemOpen
  \bibinfo {author} {L.~Frunzio}, \bibinfo {author} {A.~Wallraff}, \bibinfo {author} {D.~Schuster}, \bibinfo {author} {J.~Majer},\ and\ \bibinfo {author} {R.~Schoelkopf},\ \emph {\bibinfo {title} {Fabrication and characterization of superconducting circuit QED devices for quantum computation}},\ \href {\doibase 10.1109/TASC.2005.850084} {\bibfield  {journal} {\bibinfo  {journal} {IEEE Transactions on Applied Superconductivity}\ }\textbf {\bibinfo {volume} {15}},\ \bibinfo {pages} {860} (\bibinfo {year} {2005})}\BibitemShut {NoStop}%
\bibitem [{\citenamefont {Makhlin}\ \emph {et~al.}(2001)\citenamefont {Makhlin}, \citenamefont {Sch\"on},\ and\ \citenamefont {Shnirman}}]{RevModPhys.73.357}%
  \BibitemOpen
  \bibinfo {author} {Y.~Makhlin}, \bibinfo {author} {G.~Sch\"on},\ and\ \bibinfo {author} {A.~Shnirman},\ \emph {\bibinfo {title} {Quantum-state engineering with Josephson-junction devices}},\ \href {\doibase 10.1103/RevModPhys.73.357} {\bibfield  {journal} {\bibinfo  {journal} {Rev. Mod. Phys.}\ }\textbf {\bibinfo {volume} {73}},\ \bibinfo {pages} {357} (\bibinfo {year} {2001})}\BibitemShut {NoStop}%
\bibitem [{\citenamefont {Krantz}\ \emph {et~al.}(2019)\citenamefont {Krantz}, \citenamefont {Kjaergaard}, \citenamefont {Yan}, \citenamefont {Orlando}, \citenamefont {Gustavsson},\ and\ \citenamefont {Oliver}}]{10.1063/1.5089550}%
  \BibitemOpen
  \bibinfo {author} {P.~Krantz}, \bibinfo {author} {M.~Kjaergaard}, \bibinfo {author} {F.~Yan}, \bibinfo {author} {T.~P. Orlando}, \bibinfo {author} {S.~Gustavsson},\ and\ \bibinfo {author} {W.~D. Oliver},\ \emph {\bibinfo {title} {{A quantum engineer's guide to superconducting qubits}}},\ \href {\doibase 10.1063/1.5089550} {\bibfield  {journal} {\bibinfo  {journal} {Applied Physics Reviews}\ }\textbf {\bibinfo {volume} {6}},\ \bibinfo {pages} {021318} (\bibinfo {year} {2019})}\BibitemShut {NoStop}%
\bibitem [{\citenamefont {Blais}\ \emph {et~al.}(2020)\citenamefont {Blais}, \citenamefont {Girvin},\ and\ \citenamefont {Oliver}}]{Blais2020-oi}%
  \BibitemOpen
  \bibinfo {author} {A.~Blais}, \bibinfo {author} {S.~M. Girvin},\ and\ \bibinfo {author} {W.~D. Oliver},\ \emph {\bibinfo {title} {Quantum information processing and quantum optics with circuit quantum electrodynamics}},\ \href {\doibase https://doi.org/10.1038/s41567-020-0806-z} {\bibfield  {journal} {\bibinfo  {journal} {Nature Physics}\ }\textbf {\bibinfo {volume} {16}},\ \bibinfo {pages} {247} (\bibinfo {year} {2020})}\BibitemShut {NoStop}%
\bibitem [{\citenamefont {Devoret}\ \emph {et~al.}(2007)\citenamefont {Devoret}, \citenamefont {Girvin},\ and\ \citenamefont {Schoelkopf}}]{Devoret-howstrongcoupling}%
  \BibitemOpen
  \bibinfo {author} {M.~Devoret}, \bibinfo {author} {S.~Girvin},\ and\ \bibinfo {author} {R.~Schoelkopf},\ \emph {\bibinfo {title} {Circuit-QED: How strong can the coupling between a Josephson junction atom and a transmission line resonator be?}},\ \href {\doibase https://doi.org/10.1002/andp.200751910-1109} {\bibfield  {journal} {\bibinfo  {journal} {Annalen der Physik}\ }\textbf {\bibinfo {volume} {519}},\ \bibinfo {pages} {767} (\bibinfo {year} {2007})}\BibitemShut {NoStop}%
\bibitem [{\citenamefont {Adhikari}\ \emph {et~al.}(2013)\citenamefont {Adhikari}, \citenamefont {Hafezi},\ and\ \citenamefont {Taylor}}]{PhysRevLett.110.060503}%
  \BibitemOpen
  \bibinfo {author} {P.~Adhikari}, \bibinfo {author} {M.~Hafezi},\ and\ \bibinfo {author} {J.~M. Taylor},\ \emph {\bibinfo {title} {Nonlinear Optics Quantum Computing with Circuit QED}},\ \href {\doibase 10.1103/PhysRevLett.110.060503} {\bibfield  {journal} {\bibinfo  {journal} {Phys. Rev. Lett.}\ }\textbf {\bibinfo {volume} {110}},\ \bibinfo {pages} {060503} (\bibinfo {year} {2013})}\BibitemShut {NoStop}%
\bibitem [{\citenamefont {Deppe}\ \emph {et~al.}(2008)\citenamefont {Deppe}, \citenamefont {Mariantoni}, \citenamefont {Menzel}, \citenamefont {Marx}, \citenamefont {Saito}, \citenamefont {Kakuyanagi}, \citenamefont {Tanaka}, \citenamefont {Meno}, \citenamefont {Semba}, \citenamefont {Takayanagi}, \citenamefont {Solano},\ and\ \citenamefont {Gross}}]{Deppe2008-twoph}%
  \BibitemOpen
  \bibinfo {author} {F.~Deppe}, \bibinfo {author} {M.~Mariantoni}, \bibinfo {author} {E.~P. Menzel}, \bibinfo {author} {A.~Marx}, \bibinfo {author} {S.~Saito}, \bibinfo {author} {K.~Kakuyanagi}, \bibinfo {author} {H.~Tanaka}, \bibinfo {author} {T.~Meno}, \bibinfo {author} {K.~Semba}, \bibinfo {author} {H.~Takayanagi}, \bibinfo {author} {E.~Solano},\ and\ \bibinfo {author} {R.~Gross},\ \emph {\bibinfo {title} {Two-photon probe of the Jaynes--Cummings model and controlled symmetry breaking in circuit QED}},\ \href {\doibase 10.1038/nphys1016} {\bibfield  {journal} {\bibinfo  {journal} {Nature Physics}\ }\textbf {\bibinfo {volume} {4}},\ \bibinfo {pages} {686} (\bibinfo {year} {2008})}\BibitemShut {NoStop}%
\bibitem [{\citenamefont {Kockum}\ \emph {et~al.}(2017)\citenamefont {Kockum}, \citenamefont {Miranowicz}, \citenamefont {Macr\`{\i}}, \citenamefont {Savasta},\ and\ \citenamefont {Nori}}]{Qnonlinear-physreva-SavastaeKockum}%
  \BibitemOpen
  \bibinfo {author} {A.~F. Kockum}, \bibinfo {author} {A.~Miranowicz}, \bibinfo {author} {V.~Macr\`{\i}}, \bibinfo {author} {S.~Savasta},\ and\ \bibinfo {author} {F.~Nori},\ \emph {\bibinfo {title} {Deterministic quantum nonlinear optics with single atoms and virtual photons}},\ \href {\doibase 10.1103/PhysRevA.95.063849} {\bibfield  {journal} {\bibinfo  {journal} {Phys. Rev. A}\ }\textbf {\bibinfo {volume} {95}},\ \bibinfo {pages} {063849} (\bibinfo {year} {2017})}\BibitemShut {NoStop}%
\bibitem [{\citenamefont {Niemczyk}\ \emph {et~al.}(2010)\citenamefont {Niemczyk}, \citenamefont {Deppe}, \citenamefont {Huebl}, \citenamefont {Menzel}, \citenamefont {Hocke}, \citenamefont {Schwarz}, \citenamefont {Garcia-Ripoll}, \citenamefont {Zueco}, \citenamefont {H{\"u}mmer}, \citenamefont {Solano} \emph {et~al.}}]{Nniemczyk2010}%
  \BibitemOpen
  \bibinfo {author} {T.~Niemczyk}, \bibinfo {author} {F.~Deppe}, \bibinfo {author} {H.~Huebl}, \bibinfo {author} {E.~Menzel}, \bibinfo {author} {F.~Hocke}, \bibinfo {author} {M.~Schwarz}, \bibinfo {author} {J.~Garcia-Ripoll}, \bibinfo {author} {D.~Zueco}, \bibinfo {author} {T.~H{\"u}mmer}, \bibinfo {author} {E.~Solano} et~al.,\ \emph {\bibinfo {title} {Circuit quantum electrodynamics in the ultrastrong-coupling regime}},\ \href {\doibase https://doi.org/10.1038/nphys1730} {\bibfield  {journal} {\bibinfo  {journal} {Nature Physics}\ }\textbf {\bibinfo {volume} {6}},\ \bibinfo {pages} {772} (\bibinfo {year} {2010})}\BibitemShut {NoStop}%
\bibitem [{\citenamefont {Forn-D\'{\i}az}\ \emph {et~al.}(2010)\citenamefont {Forn-D\'{\i}az}, \citenamefont {Lisenfeld}, \citenamefont {Marcos}, \citenamefont {Garc\'{\i}a-Ripoll}, \citenamefont {Solano}, \citenamefont {Harmans},\ and\ \citenamefont {Mooij}}]{Forn2010}%
  \BibitemOpen
  \bibinfo {author} {P.~Forn-D\'{\i}az}, \bibinfo {author} {J.~Lisenfeld}, \bibinfo {author} {D.~Marcos}, \bibinfo {author} {J.~J. Garc\'{\i}a-Ripoll}, \bibinfo {author} {E.~Solano}, \bibinfo {author} {C.~J. P.~M. Harmans},\ and\ \bibinfo {author} {J.~E. Mooij},\ \emph {\bibinfo {title} {{Observation of the Bloch-Siegert Shift in a Qubit-Oscillator System in the Ultrastrong Coupling Regime}}},\ \href {\doibase 10.1103/PhysRevLett.105.237001} {\bibfield  {journal} {\bibinfo  {journal} {Phys. Rev. Lett.}\ }\textbf {\bibinfo {volume} {105}},\ \bibinfo {pages} {237001} (\bibinfo {year} {2010})}\BibitemShut {NoStop}%
\bibitem [{\citenamefont {Frisk~Kockum}\ \emph {et~al.}(2019)\citenamefont {Frisk~Kockum}, \citenamefont {Miranowicz}, \citenamefont {De~Liberato}, \citenamefont {Savasta},\ and\ \citenamefont {Nori}}]{Frisk_Kockum2019-cl}%
  \BibitemOpen
  \bibinfo {author} {A.~Frisk~Kockum}, \bibinfo {author} {A.~Miranowicz}, \bibinfo {author} {S.~De~Liberato}, \bibinfo {author} {S.~Savasta},\ and\ \bibinfo {author} {F.~Nori},\ \emph {\bibinfo {title} {Ultrastrong coupling between light and matter}},\ \href {\doibase https://doi.org/10.1038/s42254-018-0006-2} {\bibfield  {journal} {\bibinfo  {journal} {Nature Reviews Physics}\ }\textbf {\bibinfo {volume} {1}},\ \bibinfo {pages} {19} (\bibinfo {year} {2019})}\BibitemShut {NoStop}%
\bibitem [{\citenamefont {Forn-D\'{\i}az}\ \emph {et~al.}(2019)\citenamefont {Forn-D\'{\i}az}, \citenamefont {Lamata}, \citenamefont {Rico}, \citenamefont {Kono},\ and\ \citenamefont {Solano}}]{RevModPhys.91.025005}%
  \BibitemOpen
  \bibinfo {author} {P.~Forn-D\'{\i}az}, \bibinfo {author} {L.~Lamata}, \bibinfo {author} {E.~Rico}, \bibinfo {author} {J.~Kono},\ and\ \bibinfo {author} {E.~Solano},\ \emph {\bibinfo {title} {Ultrastrong coupling regimes of light-matter interaction}},\ \href {\doibase 10.1103/RevModPhys.91.025005} {\bibfield  {journal} {\bibinfo  {journal} {Rev. Mod. Phys.}\ }\textbf {\bibinfo {volume} {91}},\ \bibinfo {pages} {025005} (\bibinfo {year} {2019})}\BibitemShut {NoStop}%
\bibitem [{\citenamefont {Qin}\ \emph {et~al.}(2024)\citenamefont {Qin}, \citenamefont {Kockum}, \citenamefont {Muñoz}, \citenamefont {Miranowicz},\ and\ \citenamefont {Nori}}]{QuantumAmpl-WeiQin-Kockum-Nori}%
  \BibitemOpen
  \bibinfo {author} {W.~Qin}, \bibinfo {author} {A.~F. Kockum}, \bibinfo {author} {C.~S. Muñoz}, \bibinfo {author} {A.~Miranowicz},\ and\ \bibinfo {author} {F.~Nori},\ \emph {\bibinfo {title} {Quantum amplification and simulation of strong and ultrastrong coupling of light and matter}},\ \href {\doibase https://doi.org/10.1016/j.physrep.2024.05.003} {\bibfield  {journal} {\bibinfo  {journal} {Physics Reports}\ }\textbf {\bibinfo {volume} {1078}},\ \bibinfo {pages} {1} (\bibinfo {year} {2024})}\BibitemShut {NoStop}%
\bibitem [{\citenamefont {Forn-D{\'\i}az}\ \emph {et~al.}(2017)\citenamefont {Forn-D{\'\i}az}, \citenamefont {Garc{\'\i}a-Ripoll}, \citenamefont {Peropadre}, \citenamefont {Orgiazzi}, \citenamefont {Yurtalan}, \citenamefont {Belyansky}, \citenamefont {Wilson},\ and\ \citenamefont {Lupascu}}]{Forn-natureultrastrongcoupling-singlequbit}%
  \BibitemOpen
  \bibinfo {author} {P.~Forn-D{\'\i}az}, \bibinfo {author} {J.~J. Garc{\'\i}a-Ripoll}, \bibinfo {author} {B.~Peropadre}, \bibinfo {author} {J.-L. Orgiazzi}, \bibinfo {author} {M.~A. Yurtalan}, \bibinfo {author} {R.~Belyansky}, \bibinfo {author} {C.~M. Wilson},\ and\ \bibinfo {author} {A.~Lupascu},\ \emph {\bibinfo {title} {Ultrastrong coupling of a single artificial atom to an electromagnetic continuum in the nonperturbative regime}},\ \href {\doibase https://doi.org/10.1038/nphys3905} {\bibfield  {journal} {\bibinfo  {journal} {Nature Physics}\ }\textbf {\bibinfo {volume} {13}},\ \bibinfo {pages} {39} (\bibinfo {year} {2017})}\BibitemShut {NoStop}%
\bibitem [{\citenamefont {Bayer}\ \emph {et~al.}(2017)\citenamefont {Bayer}, \citenamefont {Pozimski}, \citenamefont {Schambeck}, \citenamefont {Schuh}, \citenamefont {Huber}, \citenamefont {Bougeard},\ and\ \citenamefont {Lange}}]{Bayer2017-xl}%
  \BibitemOpen
  \bibinfo {author} {A.~Bayer}, \bibinfo {author} {M.~Pozimski}, \bibinfo {author} {S.~Schambeck}, \bibinfo {author} {D.~Schuh}, \bibinfo {author} {R.~Huber}, \bibinfo {author} {D.~Bougeard},\ and\ \bibinfo {author} {C.~Lange},\ \emph {\bibinfo {title} {Terahertz {Light--Matter} Interaction beyond Unity Coupling Strength}},\ \href {\doibase 10.1021/acs.nanolett.7b03103} {\bibfield  {journal} {\bibinfo  {journal} {Nano Lett.}\ }\textbf {\bibinfo {volume} {17}},\ \bibinfo {pages} {6340} (\bibinfo {year} {2017})}\BibitemShut {NoStop}%
\bibitem [{\citenamefont {Yoshihara}\ \emph {et~al.}(2017{\natexlab{a}})\citenamefont {Yoshihara}, \citenamefont {Fuse}, \citenamefont {Ashhab}, \citenamefont {Kakuyanagi}, \citenamefont {Saito},\ and\ \citenamefont {Semba}}]{Yoshihara2017}%
  \BibitemOpen
  \bibinfo {author} {F.~Yoshihara}, \bibinfo {author} {T.~Fuse}, \bibinfo {author} {S.~Ashhab}, \bibinfo {author} {K.~Kakuyanagi}, \bibinfo {author} {S.~Saito},\ and\ \bibinfo {author} {K.~Semba},\ \emph {\bibinfo {title} {Superconducting qubit--oscillator circuit beyond the ultrastrong-coupling regime}},\ \href {\doibase 10.1038/nphys3906} {\bibfield  {journal} {\bibinfo  {journal} {Nature Physics}\ }\textbf {\bibinfo {volume} {13}},\ \bibinfo {pages} {44} (\bibinfo {year} {2017}{\natexlab{a}})}\BibitemShut {NoStop}%
\bibitem [{\citenamefont {Chen}\ \emph {et~al.}(2017)\citenamefont {Chen}, \citenamefont {Wang}, \citenamefont {Li}, \citenamefont {Tian}, \citenamefont {Qiu}, \citenamefont {Inomata}, \citenamefont {Yoshihara}, \citenamefont {Han}, \citenamefont {Nori}, \citenamefont {Tsai},\ and\ \citenamefont {You}}]{PhysRevA.-Singlephotonfluxqubit}%
  \BibitemOpen
  \bibinfo {author} {Z.~Chen}, \bibinfo {author} {Y.~Wang}, \bibinfo {author} {T.~Li}, \bibinfo {author} {L.~Tian}, \bibinfo {author} {Y.~Qiu}, \bibinfo {author} {K.~Inomata}, \bibinfo {author} {F.~Yoshihara}, \bibinfo {author} {S.~Han}, \bibinfo {author} {F.~Nori}, \bibinfo {author} {J.~S. Tsai},\ and\ \bibinfo {author} {J.~Q. You},\ \emph {\bibinfo {title} {Single-photon-driven high-order sideband transitions in an ultrastrongly coupled circuit-quantum-electrodynamics system}},\ \href {\doibase 10.1103/PhysRevA.96.012325} {\bibfield  {journal} {\bibinfo  {journal} {Phys. Rev. A}\ }\textbf {\bibinfo {volume} {96}},\ \bibinfo {pages} {012325} (\bibinfo {year} {2017})}\BibitemShut {NoStop}%
\bibitem [{\citenamefont {Gambino}\ \emph {et~al.}(2014)\citenamefont {Gambino}, \citenamefont {Mazzeo}, \citenamefont {Genco}, \citenamefont {Di~Stefano}, \citenamefont {Savasta}, \citenamefont {Patan{\`e}}, \citenamefont {Ballarini}, \citenamefont {Mangione}, \citenamefont {Lerario}, \citenamefont {Sanvitto},\ and\ \citenamefont {Gigli}}]{Gambino-Mazze-Savasta-ultrastrongnatural}%
  \BibitemOpen
  \bibinfo {author} {S.~Gambino}, \bibinfo {author} {M.~Mazzeo}, \bibinfo {author} {A.~Genco}, \bibinfo {author} {O.~Di~Stefano}, \bibinfo {author} {S.~Savasta}, \bibinfo {author} {S.~Patan{\`e}}, \bibinfo {author} {D.~Ballarini}, \bibinfo {author} {F.~Mangione}, \bibinfo {author} {G.~Lerario}, \bibinfo {author} {D.~Sanvitto},\ and\ \bibinfo {author} {G.~Gigli},\ \emph {\bibinfo {title} {Exploring {Light--Matter} Interaction Phenomena under Ultrastrong Coupling Regime}},\ \href {\doibase https://doi.org/10.1021/ph500266d} {\bibfield  {journal} {\bibinfo  {journal} {ACS Photonics}\ }\textbf {\bibinfo {volume} {1}},\ \bibinfo {pages} {1042} (\bibinfo {year} {2014})}\BibitemShut {NoStop}%
\bibitem [{\citenamefont {Forn-D{\'i}az}\ \emph {et~al.}(2016)\citenamefont {Forn-D{\'i}az}, \citenamefont {Romero}, \citenamefont {Harmans}, \citenamefont {Solano},\ and\ \citenamefont {Mooij}}]{Forn-Díaz2016-brokensrule}%
  \BibitemOpen
  \bibinfo {author} {P.~Forn-D{\'i}az}, \bibinfo {author} {G.~Romero}, \bibinfo {author} {C.~J. P.~M. Harmans}, \bibinfo {author} {E.~Solano},\ and\ \bibinfo {author} {J.~E. Mooij},\ \emph {\bibinfo {title} {Broken selection rule in the quantum Rabi model}},\ \href {\doibase 10.1038/srep26720} {\bibfield  {journal} {\bibinfo  {journal} {Scientific Reports}\ }\textbf {\bibinfo {volume} {6}},\ \bibinfo {pages} {26720} (\bibinfo {year} {2016})}\BibitemShut {NoStop}%
\bibitem [{\citenamefont {Bernardis}\ \emph {et~al.}(2024)\citenamefont {Bernardis}, \citenamefont {Mercurio},\ and\ \citenamefont {Liberato}}]{DeBernardis:24_Alberto_Review}%
  \BibitemOpen
  \bibinfo {author} {D.~D. Bernardis}, \bibinfo {author} {A.~Mercurio},\ and\ \bibinfo {author} {S.~D. Liberato},\ \emph {\bibinfo {title} {Tutorial on nonperturbative cavity quantum electrodynamics: is the Jaynes-Cummings model still relevant?}},\ \href {\doibase 10.1364/JOSAB.522786} {\bibfield  {journal} {\bibinfo  {journal} {J. Opt. Soc. Am. B}\ }\textbf {\bibinfo {volume} {41}},\ \bibinfo {pages} {C206} (\bibinfo {year} {2024})}\BibitemShut {NoStop}%
\bibitem [{\citenamefont {Le~Boité}(2020)}]{Lwboitè-reviewultra}%
  \BibitemOpen
  \bibinfo {author} {A.~Le~Boité},\ \emph {\bibinfo {title} {Theoretical Methods for Ultrastrong Light–Matter Interactions}},\ \href {\doibase https://doi.org/10.1002/qute.201900140} {\bibfield  {journal} {\bibinfo  {journal} {Advanced Quantum Technologies}\ }\textbf {\bibinfo {volume} {3}},\ \bibinfo {pages} {1900140} (\bibinfo {year} {2020})}\BibitemShut {NoStop}%
\bibitem [{\citenamefont {Wang}\ \emph {et~al.}(2024)\citenamefont {Wang}, \citenamefont {Mercurio}, \citenamefont {Ridolfo}, \citenamefont {Wang}, \citenamefont {Chen}, \citenamefont {Li}, \citenamefont {Nori}, \citenamefont {Savasta},\ and\ \citenamefont {You}}]{wang2024strong}%
  \BibitemOpen
  \bibinfo {author} {S.-P. Wang}, \bibinfo {author} {A.~Mercurio}, \bibinfo {author} {A.~Ridolfo}, \bibinfo {author} {Y.~Wang}, \bibinfo {author} {M.~Chen}, \bibinfo {author} {T.~Li}, \bibinfo {author} {F.~Nori}, \bibinfo {author} {S.~Savasta},\ and\ \bibinfo {author} {J.~Q. You},\ \href@noop {} {\emph {\bibinfo {title} {Strong coupling between a single photon and a photon pair}}} (\bibinfo {year} {2024}),\ \Eprint {http://arxiv.org/abs/2401.02738} {arXiv:2401.02738 [quant-ph]} \BibitemShut {NoStop}%
\bibitem [{\citenamefont {Garziano}\ \emph {et~al.}(2016)\citenamefont {Garziano}, \citenamefont {Macr\`{\i}}, \citenamefont {Stassi}, \citenamefont {Di~Stefano}, \citenamefont {Nori},\ and\ \citenamefont {Savasta}}]{PhysRevLet-onephtwoatoms}%
  \BibitemOpen
  \bibinfo {author} {L.~Garziano}, \bibinfo {author} {V.~Macr\`{\i}}, \bibinfo {author} {R.~Stassi}, \bibinfo {author} {O.~Di~Stefano}, \bibinfo {author} {F.~Nori},\ and\ \bibinfo {author} {S.~Savasta},\ \emph {\bibinfo {title} {One Photon Can Simultaneously Excite Two or More Atoms}},\ \href {\doibase 10.1103/PhysRevLett.117.043601} {\bibfield  {journal} {\bibinfo  {journal} {Phys. Rev. Lett.}\ }\textbf {\bibinfo {volume} {117}},\ \bibinfo {pages} {043601} (\bibinfo {year} {2016})}\BibitemShut {NoStop}%
\bibitem [{\citenamefont {Giannelli}\ \emph {et~al.}(2024)\citenamefont {Giannelli}, \citenamefont {Paladino}, \citenamefont {Grajcar}, \citenamefont {Paraoanu},\ and\ \citenamefont {Falci}}]{VirtualFalci}%
  \BibitemOpen
  \bibinfo {author} {L.~Giannelli}, \bibinfo {author} {E.~Paladino}, \bibinfo {author} {M.~Grajcar}, \bibinfo {author} {G.~S. Paraoanu},\ and\ \bibinfo {author} {G.~Falci},\ \emph {\bibinfo {title} {Detecting virtual photons in ultrastrongly coupled superconducting quantum circuits}},\ \href {\doibase 10.1103/PhysRevResearch.6.013008} {\bibfield  {journal} {\bibinfo  {journal} {Phys. Rev. Res.}\ }\textbf {\bibinfo {volume} {6}},\ \bibinfo {pages} {013008} (\bibinfo {year} {2024})}\BibitemShut {NoStop}%
\bibitem [{\citenamefont {Tomonaga}\ \emph {et~al.}(2024)\citenamefont {Tomonaga}, \citenamefont {Stassi}, \citenamefont {Mukai}, \citenamefont {Nori}, \citenamefont {Yoshihara},\ and\ \citenamefont {Tsai}}]{tomonaga2024photonsimultaneouslyexcitesatoms}%
  \BibitemOpen
  \bibinfo {author} {A.~Tomonaga}, \bibinfo {author} {R.~Stassi}, \bibinfo {author} {H.~Mukai}, \bibinfo {author} {F.~Nori}, \bibinfo {author} {F.~Yoshihara},\ and\ \bibinfo {author} {J.-S. Tsai},\ \href@noop {} {\emph {\bibinfo {title} {One photon simultaneously excites two atoms in a ultrastrongly coupled light-matter system}}} (\bibinfo {year} {2024}),\ \Eprint {http://arxiv.org/abs/2307.15437} {arXiv:2307.15437 [quant-ph]} \BibitemShut {NoStop}%
\bibitem [{\citenamefont {Choi}(2020)}]{Exoticquantumsate-cqed}%
  \BibitemOpen
  \bibinfo {author} {M.-S. Choi},\ \emph {\bibinfo {title} {Exotic Quantum States of Circuit Quantum Electrodynamics in the Ultra-Strong Coupling Regime}},\ \href {\doibase https://doi.org/10.1002/qute.202000085} {\bibfield  {journal} {\bibinfo  {journal} {Advanced Quantum Technologies}\ }\textbf {\bibinfo {volume} {3}},\ \bibinfo {pages} {2000085} (\bibinfo {year} {2020})}\BibitemShut {NoStop}%
\bibitem [{\citenamefont {Ridolfo}\ \emph {et~al.}(2012)\citenamefont {Ridolfo}, \citenamefont {Leib}, \citenamefont {Savasta},\ and\ \citenamefont {Hartmann}}]{Photon-Blockade-SavastaEridolfo}%
  \BibitemOpen
  \bibinfo {author} {A.~Ridolfo}, \bibinfo {author} {M.~Leib}, \bibinfo {author} {S.~Savasta},\ and\ \bibinfo {author} {M.~J. Hartmann},\ \emph {\bibinfo {title} {Photon Blockade in the Ultrastrong Coupling Regime}},\ \href {\doibase 10.1103/PhysRevLett.109.193602} {\bibfield  {journal} {\bibinfo  {journal} {Phys. Rev. Lett.}\ }\textbf {\bibinfo {volume} {109}},\ \bibinfo {pages} {193602} (\bibinfo {year} {2012})}\BibitemShut {NoStop}%
\bibitem [{\citenamefont {Wang}\ \emph {et~al.}(2023)\citenamefont {Wang}, \citenamefont {Ridolfo}, \citenamefont {Li}, \citenamefont {Savasta}, \citenamefont {Nori}, \citenamefont {Nakamura},\ and\ \citenamefont {You}}]{Wang2023}%
  \BibitemOpen
  \bibinfo {author} {S.-P. Wang}, \bibinfo {author} {A.~Ridolfo}, \bibinfo {author} {T.~Li}, \bibinfo {author} {S.~Savasta}, \bibinfo {author} {F.~Nori}, \bibinfo {author} {Y.~Nakamura},\ and\ \bibinfo {author} {J.~Q. You},\ \emph {\bibinfo {title} {Probing the symmetry breaking of a light--matter system by an ancillary qubit}},\ \href {\doibase 10.1038/s41467-023-40097-0} {\bibfield  {journal} {\bibinfo  {journal} {Nature Communications}\ }\textbf {\bibinfo {volume} {14}},\ \bibinfo {pages} {4397} (\bibinfo {year} {2023})}\BibitemShut {NoStop}%
\bibitem [{\citenamefont {Yoshihara}\ \emph {et~al.}(2017{\natexlab{b}})\citenamefont {Yoshihara}, \citenamefont {Fuse}, \citenamefont {Ashhab}, \citenamefont {Kakuyanagi}, \citenamefont {Saito},\ and\ \citenamefont {Semba}}]{Phys.Rev.Semba}%
  \BibitemOpen
  \bibinfo {author} {F.~Yoshihara}, \bibinfo {author} {T.~Fuse}, \bibinfo {author} {S.~Ashhab}, \bibinfo {author} {K.~Kakuyanagi}, \bibinfo {author} {S.~Saito},\ and\ \bibinfo {author} {K.~Semba},\ \emph {\bibinfo {title} {Characteristic spectra of circuit quantum electrodynamics systems from the ultrastrong- to the deep-strong-coupling regime}},\ \href {\doibase 10.1103/PhysRevA.95.053824} {\bibfield  {journal} {\bibinfo  {journal} {Phys. Rev. A}\ }\textbf {\bibinfo {volume} {95}},\ \bibinfo {pages} {053824} (\bibinfo {year} {2017}{\natexlab{b}})}\BibitemShut {NoStop}%
\bibitem [{\citenamefont {Chiorescu}\ \emph {et~al.}(2004)\citenamefont {Chiorescu}, \citenamefont {Bertet}, \citenamefont {Semba}, \citenamefont {Nakamura}, \citenamefont {Harmans},\ and\ \citenamefont {Mooij}}]{Chiorescu2004}%
  \BibitemOpen
  \bibinfo {author} {I.~Chiorescu}, \bibinfo {author} {P.~Bertet}, \bibinfo {author} {K.~Semba}, \bibinfo {author} {Y.~Nakamura}, \bibinfo {author} {C.~J. P.~M. Harmans},\ and\ \bibinfo {author} {J.~E. Mooij},\ \emph {\bibinfo {title} {Coherent dynamics of a flux qubit coupled to a harmonic oscillator}},\ \href {\doibase 10.1038/nature02831} {\bibfield  {journal} {\bibinfo  {journal} {Nature}\ }\textbf {\bibinfo {volume} {431}},\ \bibinfo {pages} {159} (\bibinfo {year} {2004})}\BibitemShut {NoStop}%
\bibitem [{\citenamefont {Yan}\ \emph {et~al.}(2016)\citenamefont {Yan}, \citenamefont {Gustavsson}, \citenamefont {Kamal}, \citenamefont {Birenbaum}, \citenamefont {Sears}, \citenamefont {Hover}, \citenamefont {Gudmundsen}, \citenamefont {Rosenberg}, \citenamefont {Samach}, \citenamefont {Weber}, \citenamefont {Yoder}, \citenamefont {Orlando}, \citenamefont {Clarke}, \citenamefont {Kerman},\ and\ \citenamefont {Oliver}}]{Yan2016-or}%
  \BibitemOpen
  \bibinfo {author} {F.~Yan}, \bibinfo {author} {S.~Gustavsson}, \bibinfo {author} {A.~Kamal}, \bibinfo {author} {J.~Birenbaum}, \bibinfo {author} {A.~P. Sears}, \bibinfo {author} {D.~Hover}, \bibinfo {author} {T.~J. Gudmundsen}, \bibinfo {author} {D.~Rosenberg}, \bibinfo {author} {G.~Samach}, \bibinfo {author} {S.~Weber}, \bibinfo {author} {J.~L. Yoder}, \bibinfo {author} {T.~P. Orlando}, \bibinfo {author} {J.~Clarke}, \bibinfo {author} {A.~J. Kerman},\ and\ \bibinfo {author} {W.~D. Oliver},\ \emph {\bibinfo {title} {The flux qubit revisited to enhance coherence and reproducibility}},\ \href {\doibase https://doi.org/10.1038/ncomms12964} {\bibfield  {journal} {\bibinfo  {journal} {Nature Communications}\ }\textbf {\bibinfo {volume} {7}},\ \bibinfo {pages} {12964} (\bibinfo {year} {2016})}\BibitemShut {NoStop}%
\bibitem [{\citenamefont {Saito}\ \emph {et~al.}(2006)\citenamefont {Saito}, \citenamefont {Meno}, \citenamefont {Ueda}, \citenamefont {Tanaka}, \citenamefont {Semba},\ and\ \citenamefont {Takayanagi}}]{PhysRevLett-Semba-Saito-fluxcontrol}%
  \BibitemOpen
  \bibinfo {author} {S.~Saito}, \bibinfo {author} {T.~Meno}, \bibinfo {author} {M.~Ueda}, \bibinfo {author} {H.~Tanaka}, \bibinfo {author} {K.~Semba},\ and\ \bibinfo {author} {H.~Takayanagi},\ \emph {\bibinfo {title} {Parametric Control of a Superconducting Flux Qubit}},\ \href {\doibase 10.1103/PhysRevLett.96.107001} {\bibfield  {journal} {\bibinfo  {journal} {Phys. Rev. Lett.}\ }\textbf {\bibinfo {volume} {96}},\ \bibinfo {pages} {107001} (\bibinfo {year} {2006})}\BibitemShut {NoStop}%
\bibitem [{\citenamefont {Yoshihara}\ \emph {et~al.}(2022)\citenamefont {Yoshihara}, \citenamefont {Ashhab}, \citenamefont {Fuse}, \citenamefont {Bamba},\ and\ \citenamefont {Semba}}]{Yoshihara2022}%
  \BibitemOpen
  \bibinfo {author} {F.~Yoshihara}, \bibinfo {author} {S.~Ashhab}, \bibinfo {author} {T.~Fuse}, \bibinfo {author} {M.~Bamba},\ and\ \bibinfo {author} {K.~Semba},\ \emph {\bibinfo {title} {Hamiltonian of a flux qubit-LC oscillator circuit in the deep--strong-coupling regime}},\ \href {\doibase 10.1038/s41598-022-10203-1} {\bibfield  {journal} {\bibinfo  {journal} {Scientific Reports}\ }\textbf {\bibinfo {volume} {12}},\ \bibinfo {pages} {6764} (\bibinfo {year} {2022})}\BibitemShut {NoStop}%
\bibitem [{\citenamefont {Manucharyan}\ \emph {et~al.}(2017)\citenamefont {Manucharyan}, \citenamefont {Baksic},\ and\ \citenamefont {Ciuti}}]{Manucharyan_2017}%
  \BibitemOpen
  \bibinfo {author} {V.~E. Manucharyan}, \bibinfo {author} {A.~Baksic},\ and\ \bibinfo {author} {C.~Ciuti},\ \emph {\bibinfo {title} {Resilience of the quantum Rabi model in circuit QED}},\ \href {\doibase 10.1088/1751-8121/aa6fbc} {\bibfield  {journal} {\bibinfo  {journal} {Journal of Physics A: Mathematical and Theoretical}\ }\textbf {\bibinfo {volume} {50}},\ \bibinfo {pages} {294001} (\bibinfo {year} {2017})}\BibitemShut {NoStop}%
\bibitem [{\citenamefont {Savasta}\ \emph {et~al.}(2021)\citenamefont {Savasta}, \citenamefont {Di~Stefano}, \citenamefont {Settineri}, \citenamefont {Zueco}, \citenamefont {Hughes},\ and\ \citenamefont {Nori}}]{PhysRevA.103.053703}%
  \BibitemOpen
  \bibinfo {author} {S.~Savasta}, \bibinfo {author} {O.~Di~Stefano}, \bibinfo {author} {A.~Settineri}, \bibinfo {author} {D.~Zueco}, \bibinfo {author} {S.~Hughes},\ and\ \bibinfo {author} {F.~Nori},\ \emph {\bibinfo {title} {Gauge principle and gauge invariance in two-level systems}},\ \href {\doibase 10.1103/PhysRevA.103.053703} {\bibfield  {journal} {\bibinfo  {journal} {Phys. Rev. A}\ }\textbf {\bibinfo {volume} {103}},\ \bibinfo {pages} {053703} (\bibinfo {year} {2021})}\BibitemShut {NoStop}%
\bibitem [{\citenamefont {Settineri}\ \emph {et~al.}(2018)\citenamefont {Settineri}, \citenamefont {Macr\'{\i}}, \citenamefont {Ridolfo}, \citenamefont {Di~Stefano}, \citenamefont {Kockum}, \citenamefont {Nori},\ and\ \citenamefont {Savasta}}]{PhysRevA.98.053834}%
  \BibitemOpen
  \bibinfo {author} {A.~Settineri}, \bibinfo {author} {V.~Macr\'{\i}}, \bibinfo {author} {A.~Ridolfo}, \bibinfo {author} {O.~Di~Stefano}, \bibinfo {author} {A.~F. Kockum}, \bibinfo {author} {F.~Nori},\ and\ \bibinfo {author} {S.~Savasta},\ \emph {\bibinfo {title} {Dissipation and thermal noise in hybrid quantum systems in the ultrastrong-coupling regime}},\ \href {\doibase 10.1103/PhysRevA.98.053834} {\bibfield  {journal} {\bibinfo  {journal} {Phys. Rev. A}\ }\textbf {\bibinfo {volume} {98}},\ \bibinfo {pages} {053834} (\bibinfo {year} {2018})}\BibitemShut {NoStop}%
\bibitem [{\citenamefont {Gardiner}\ and\ \citenamefont {Zoller}(2004)}]{gardiner2004quantum}%
  \BibitemOpen
  \bibinfo {author} {C.~Gardiner}\ and\ \bibinfo {author} {P.~Zoller},\ \href@noop {} {\emph {\bibinfo {title} {Quantum noise: a handbook of Markovian and non-Markovian quantum stochastic methods with applications to quantum optics}}}\ (\bibinfo  {publisher} {Springer Science \& Business Media},\ \bibinfo {year} {2004})\BibitemShut {NoStop}%
\bibitem [{\citenamefont {Mercurio}\ \emph {et~al.}(2022)\citenamefont {Mercurio}, \citenamefont {Macr\`{\i}}, \citenamefont {Gustin}, \citenamefont {Hughes}, \citenamefont {Savasta},\ and\ \citenamefont {Nori}}]{PhysRevResearch.4.023048}%
  \BibitemOpen
  \bibinfo {author} {A.~Mercurio}, \bibinfo {author} {V.~Macr\`{\i}}, \bibinfo {author} {C.~Gustin}, \bibinfo {author} {S.~Hughes}, \bibinfo {author} {S.~Savasta},\ and\ \bibinfo {author} {F.~Nori},\ \emph {\bibinfo {title} {Regimes of cavity QED under incoherent excitation: From weak to deep strong coupling}},\ \href {\doibase 10.1103/PhysRevResearch.4.023048} {\bibfield  {journal} {\bibinfo  {journal} {Phys. Rev. Res.}\ }\textbf {\bibinfo {volume} {4}},\ \bibinfo {pages} {023048} (\bibinfo {year} {2022})}\BibitemShut {NoStop}%
\bibitem [{\citenamefont {Walls}\ and\ \citenamefont {Milburn}(2008)}]{Walls2008}%
  \BibitemOpen
  \bibinfo {author} {D.~Walls}\ and\ \bibinfo {author} {G.~J. Milburn},\ \emph {\bibinfo {title} {Input--Output Formulation of Optical Cavities}},\ in\ \href {\doibase 10.1007/978-3-540-28574-8_7} {\emph {\bibinfo {booktitle} {Quantum Optics}}},\ \bibinfo {editor} {edited by\ \bibinfo {editor} {D.~Walls}\ and\ \bibinfo {editor} {G.~J. Milburn}}\ (\bibinfo  {publisher} {Springer Berlin Heidelberg},\ \bibinfo {address} {Berlin, Heidelberg},\ \bibinfo {year} {2008})\ pp.\ \bibinfo {pages} {127--141}\BibitemShut {NoStop}%
\bibitem [{\citenamefont {Macr\`{\i}}\ \emph {et~al.}(2022)\citenamefont {Macr\`{\i}}, \citenamefont {Mercurio}, \citenamefont {Nori}, \citenamefont {Savasta},\ and\ \citenamefont {S\'anchez Mu\~noz}}]{Raman-Alberto}%
  \BibitemOpen
  \bibinfo {author} {V.~Macr\`{\i}}, \bibinfo {author} {A.~Mercurio}, \bibinfo {author} {F.~Nori}, \bibinfo {author} {S.~Savasta},\ and\ \bibinfo {author} {C.~S\'anchez Mu\~noz},\ \emph {\bibinfo {title} {Spontaneous Scattering of Raman Photons from Cavity-QED Systems in the Ultrastrong Coupling Regime}},\ \href {\doibase 10.1103/PhysRevLett.129.273602} {\bibfield  {journal} {\bibinfo  {journal} {Phys. Rev. Lett.}\ }\textbf {\bibinfo {volume} {129}},\ \bibinfo {pages} {273602} (\bibinfo {year} {2022})}\BibitemShut {NoStop}%
\bibitem [{\citenamefont {Salmon}\ \emph {et~al.}(2022)\citenamefont {Salmon}, \citenamefont {Gustin}, \citenamefont {Settineri}, \citenamefont {Stefano}, \citenamefont {Zueco}, \citenamefont {Savasta}, \citenamefont {Nori},\ and\ \citenamefont {Hughes}}]{SalmonGustinSettineriDiStefanoZuecoSavastaNoriHughees-guagefreedomspectra}%
  \BibitemOpen
  \bibinfo {author} {W.~Salmon}, \bibinfo {author} {C.~Gustin}, \bibinfo {author} {A.~Settineri}, \bibinfo {author} {O.~D. Stefano}, \bibinfo {author} {D.~Zueco}, \bibinfo {author} {S.~Savasta}, \bibinfo {author} {F.~Nori},\ and\ \bibinfo {author} {S.~Hughes},\ \emph {\bibinfo {title} {{Gauge-independent emission spectra and quantum correlations in the ultrastrong coupling regime of open system cavity-QED}}},\ \href {\doibase doi:10.1515/nanoph-2021-0718} {\bibfield  {journal} {\bibinfo  {journal} {Nanophotonics}\ }\textbf {\bibinfo {volume} {11}},\ \bibinfo {pages} {1573} (\bibinfo {year} {2022})}\BibitemShut {NoStop}%
\bibitem [{\citenamefont {Mercurio}\ \emph {et~al.}()\citenamefont {Mercurio}, \citenamefont {Gravina},\ and\ \citenamefont {Huang.}}]{quantumtoolboxjl}%
  \BibitemOpen
  \bibinfo {author} {A.~Mercurio}, \bibinfo {author} {L.~Gravina},\ and\ \bibinfo {author} {Y.-T. Huang.},\ \href {https://qutip.org/QuantumToolbox.jl/stable/} {\emph {\bibinfo {title} {QuantumToolbox.jl: A Julia package for quantum information and computing}}},\ \bibinfo {note} {version v0.11.4}\BibitemShut {NoStop}%
\bibitem [{\citenamefont {Johansson}\ \emph {et~al.}(2012)\citenamefont {Johansson}, \citenamefont {Nation},\ and\ \citenamefont {Nori}}]{JOHANSSON20121760}%
  \BibitemOpen
  \bibinfo {author} {J.~Johansson}, \bibinfo {author} {P.~Nation},\ and\ \bibinfo {author} {F.~Nori},\ \emph {\bibinfo {title} {QuTiP: An open-source Python framework for the dynamics of open quantum systems}},\ \href {\doibase https://doi.org/10.1016/j.cpc.2012.02.021} {\bibfield  {journal} {\bibinfo  {journal} {Computer Physics Communications}\ }\textbf {\bibinfo {volume} {183}},\ \bibinfo {pages} {1760} (\bibinfo {year} {2012})}\BibitemShut {NoStop}%
\bibitem [{\citenamefont {Johansson}\ \emph {et~al.}(2013)\citenamefont {Johansson}, \citenamefont {Nation},\ and\ \citenamefont {Nori}}]{JOHANSSON20131234}%
  \BibitemOpen
  \bibinfo {author} {J.~Johansson}, \bibinfo {author} {P.~Nation},\ and\ \bibinfo {author} {F.~Nori},\ \emph {\bibinfo {title} {QuTiP 2: A Python framework for the dynamics of open quantum systems}},\ \href {\doibase https://doi.org/10.1016/j.cpc.2012.11.019} {\bibfield  {journal} {\bibinfo  {journal} {Computer Physics Communications}\ }\textbf {\bibinfo {volume} {184}},\ \bibinfo {pages} {1234} (\bibinfo {year} {2013})}\BibitemShut {NoStop}%
\bibitem [{\citenamefont {Beaudoin}\ \emph {et~al.}(2011)\citenamefont {Beaudoin}, \citenamefont {Gambetta},\ and\ \citenamefont {Blais}}]{PhysRevA.84.043832}%
  \BibitemOpen
  \bibinfo {author} {F.~Beaudoin}, \bibinfo {author} {J.~M. Gambetta},\ and\ \bibinfo {author} {A.~Blais},\ \emph {\bibinfo {title} {Dissipation and ultrastrong coupling in circuit QED}},\ \href {\doibase 10.1103/PhysRevA.84.043832} {\bibfield  {journal} {\bibinfo  {journal} {Phys. Rev. A}\ }\textbf {\bibinfo {volume} {84}},\ \bibinfo {pages} {043832} (\bibinfo {year} {2011})}\BibitemShut {NoStop}%
\bibitem [{\citenamefont {De~Liberato}(2014)}]{Decoupling-deliberato}%
  \BibitemOpen
  \bibinfo {author} {S.~De~Liberato},\ \emph {\bibinfo {title} {Light-Matter Decoupling in the Deep Strong Coupling Regime: The Breakdown of the Purcell Effect}},\ \href {\doibase 10.1103/PhysRevLett.112.016401} {\bibfield  {journal} {\bibinfo  {journal} {Phys. Rev. Lett.}\ }\textbf {\bibinfo {volume} {112}},\ \bibinfo {pages} {016401} (\bibinfo {year} {2014})}\BibitemShut {NoStop}%
\bibitem [{\citenamefont {Salado-Mejía}\ \emph {et~al.}(2021)\citenamefont {Salado-Mejía}, \citenamefont {Román-Ancheyta}, \citenamefont {Soto-Eguibar},\ and\ \citenamefont {Moya-Cessa}}]{Salado-Mejía_2021_thermodynamicsultrastrong}%
  \BibitemOpen
  \bibinfo {author} {M.~Salado-Mejía}, \bibinfo {author} {R.~Román-Ancheyta}, \bibinfo {author} {F.~Soto-Eguibar},\ and\ \bibinfo {author} {H.~M. Moya-Cessa},\ \emph {\bibinfo {title} {Spectroscopy and critical quantum thermometry in the ultrastrong coupling regime}},\ \href {\doibase 10.1088/2058-9565/abdca5} {\bibfield  {journal} {\bibinfo  {journal} {Quantum Science and Technology}\ }\textbf {\bibinfo {volume} {6}},\ \bibinfo {pages} {025010} (\bibinfo {year} {2021})}\BibitemShut {NoStop}%
\bibitem [{\citenamefont {Babiker}\ \emph {et~al.}(1983)\citenamefont {Babiker}, \citenamefont {Loudon},\ and\ \citenamefont {Series}}]{Babiker-Loudon-Gauge}%
  \BibitemOpen
  \bibinfo {author} {M.~Babiker}, \bibinfo {author} {R.~Loudon},\ and\ \bibinfo {author} {G.~W. Series},\ \emph {\bibinfo {title} {Derivation of the Power-Zienau-Woolley Hamiltonian in quantum electrodynamics by gauge transformation}},\ \href {\doibase 10.1098/rspa.1983.0022} {\bibfield  {journal} {\bibinfo  {journal} {Proceedings of the Royal Society of London. A. Mathematical and Physical Sciences}\ }\textbf {\bibinfo {volume} {385}},\ \bibinfo {pages} {439} (\bibinfo {year} {1983})}\BibitemShut {NoStop}%
\bibitem [{\citenamefont {Cohen-Tannoudji}\ \emph {et~al.}(1997)\citenamefont {Cohen-Tannoudji}, \citenamefont {Dupont-Roc},\ and\ \citenamefont {Grynberg}}]{CohenTannoudji1997}%
  \BibitemOpen
  \bibinfo {author} {C.~Cohen-Tannoudji}, \bibinfo {author} {J.~Dupont-Roc},\ and\ \bibinfo {author} {G.~Grynberg},\ \href@noop {} {\emph {\bibinfo {title} {Photons and Atoms: Introduction to Quantum Electrodynamics}}}\ (\bibinfo  {publisher} {Wiley-VCH},\ \bibinfo {year} {1997})\BibitemShut {NoStop}%
\bibitem [{\citenamefont {Di~Stefano}\ \emph {et~al.}(2019)\citenamefont {Di~Stefano}, \citenamefont {Settineri}, \citenamefont {Macr{\`i}}, \citenamefont {Garziano}, \citenamefont {Stassi}, \citenamefont {Savasta},\ and\ \citenamefont {Nori}}]{DiStefano2019}%
  \BibitemOpen
  \bibinfo {author} {O.~Di~Stefano}, \bibinfo {author} {A.~Settineri}, \bibinfo {author} {V.~Macr{\`i}}, \bibinfo {author} {L.~Garziano}, \bibinfo {author} {R.~Stassi}, \bibinfo {author} {S.~Savasta},\ and\ \bibinfo {author} {F.~Nori},\ \emph {\bibinfo {title} {Resolution of gauge ambiguities in ultrastrong-coupling cavity quantum electrodynamics}},\ \href {\doibase 10.1038/s41567-019-0534-4} {\bibfield  {journal} {\bibinfo  {journal} {Nature Physics}\ }\textbf {\bibinfo {volume} {15}},\ \bibinfo {pages} {803} (\bibinfo {year} {2019})}\BibitemShut {NoStop}%
\bibitem [{\citenamefont {Gustin}\ \emph {et~al.}(2023)\citenamefont {Gustin}, \citenamefont {Franke},\ and\ \citenamefont {Hughes}}]{Gauge-inviariant-Hughes}%
  \BibitemOpen
  \bibinfo {author} {C.~Gustin}, \bibinfo {author} {S.~Franke},\ and\ \bibinfo {author} {S.~Hughes},\ \emph {\bibinfo {title} {Gauge-invariant theory of truncated quantum light-matter interactions in arbitrary media}},\ \href {\doibase 10.1103/PhysRevA.107.013722} {\bibfield  {journal} {\bibinfo  {journal} {Phys. Rev. A}\ }\textbf {\bibinfo {volume} {107}},\ \bibinfo {pages} {013722} (\bibinfo {year} {2023})}\BibitemShut {NoStop}%
\bibitem [{\citenamefont {Settineri}\ \emph {et~al.}(2021)\citenamefont {Settineri}, \citenamefont {Di~Stefano}, \citenamefont {Zueco}, \citenamefont {Hughes}, \citenamefont {Savasta},\ and\ \citenamefont {Nori}}]{Settineri_2021}%
  \BibitemOpen
  \bibinfo {author} {A.~Settineri}, \bibinfo {author} {O.~Di~Stefano}, \bibinfo {author} {D.~Zueco}, \bibinfo {author} {S.~Hughes}, \bibinfo {author} {S.~Savasta},\ and\ \bibinfo {author} {F.~Nori},\ \emph {\bibinfo {title} {Gauge freedom, quantum measurements, and time-dependent interactions in cavity QED}},\ \href {\doibase 10.1103/PhysRevResearch.3.023079} {\bibfield  {journal} {\bibinfo  {journal} {Phys. Rev. Res.}\ }\textbf {\bibinfo {volume} {3}},\ \bibinfo {pages} {023079} (\bibinfo {year} {2021})}\BibitemShut {NoStop}%
\bibitem [{\citenamefont {Glauber}(1963)}]{GlauberPhysRev}%
  \BibitemOpen
  \bibinfo {author} {R.~J. Glauber},\ \emph {\bibinfo {title} {The Quantum Theory of Optical Coherence}},\ \href {\doibase 10.1103/PhysRev.130.2529} {\bibfield  {journal} {\bibinfo  {journal} {Phys. Rev.}\ }\textbf {\bibinfo {volume} {130}},\ \bibinfo {pages} {2529} (\bibinfo {year} {1963})}\BibitemShut {NoStop}%
\bibitem [{\citenamefont {Mercurio}\ \emph {et~al.}(2023)\citenamefont {Mercurio}, \citenamefont {Abo}, \citenamefont {Mauceri}, \citenamefont {Russo}, \citenamefont {Macr\`{\i}}, \citenamefont {Miranowicz}, \citenamefont {Savasta},\ and\ \citenamefont {Di~Stefano}}]{PureDeph-Alberto}%
  \BibitemOpen
  \bibinfo {author} {A.~Mercurio}, \bibinfo {author} {S.~Abo}, \bibinfo {author} {F.~Mauceri}, \bibinfo {author} {E.~Russo}, \bibinfo {author} {V.~Macr\`{\i}}, \bibinfo {author} {A.~Miranowicz}, \bibinfo {author} {S.~Savasta},\ and\ \bibinfo {author} {O.~Di~Stefano},\ \emph {\bibinfo {title} {Pure Dephasing of Light-Matter Systems in the Ultrastrong and Deep-Strong Coupling Regimes}},\ \href {\doibase 10.1103/PhysRevLett.130.123601} {\bibfield  {journal} {\bibinfo  {journal} {Phys. Rev. Lett.}\ }\textbf {\bibinfo {volume} {130}},\ \bibinfo {pages} {123601} (\bibinfo {year} {2023})}\BibitemShut {NoStop}%
\bibitem [{\citenamefont {Economou}(2006)}]{economou2006green}%
  \BibitemOpen
  \bibinfo {author} {E.~N. Economou},\ \href@noop {} {\emph {\bibinfo {title} {Green's functions in quantum physics}}},\ Vol.~\bibinfo {volume} {7}\ (\bibinfo  {publisher} {Springer Science \& Business Media},\ \bibinfo {year} {2006})\BibitemShut {NoStop}%
\bibitem [{\citenamefont {Yurke}\ and\ \citenamefont {Denker}(1984)}]{Quantumnet}%
  \BibitemOpen
  \bibinfo {author} {B.~Yurke}\ and\ \bibinfo {author} {J.~S. Denker},\ \emph {\bibinfo {title} {Quantum network theory}},\ \href {\doibase 10.1103/PhysRevA.29.1419} {\bibfield  {journal} {\bibinfo  {journal} {Phys. Rev. A}\ }\textbf {\bibinfo {volume} {29}},\ \bibinfo {pages} {1419} (\bibinfo {year} {1984})}\BibitemShut {NoStop}%
\end{thebibliography}%

\end{document}